\documentclass[aps,prx,reprint,superscriptaddress]{revtex4-2}

\usepackage{amsmath,amssymb,amsfonts}
\usepackage{physics}
\usepackage{bbm}
\usepackage{bm}
\usepackage{hyperref}
\usepackage{mathtools}
\usepackage{enumitem}
\usepackage[utf8]{inputenc}
\usepackage{graphicx,color}
\usepackage{esint}
\usepackage{empheq}
\usepackage[dvipsnames]{xcolor}

\hypersetup{
    colorlinks=true,
    linkcolor=blue,
    citecolor=blue,
    urlcolor=blue
}

\begin{document}

\title[Article Title]{Physical Adjoint Learning in Wave Systems with Asymmetric Couplings}

\author{Lucas J. Fernández-Alcázar}
\email{lfernandez@exa.unne.edu.ar}
\affiliation{Wave Transport in Complex Systems Lab, Department of Physics, Wesleyan University, Middletown, Connecticut 06459, USA}
\affiliation{Institute for Modeling and Innovative Technology, IMIT (CONICET - UNNE), Corrientes, Argentina}
\affiliation{Physics Department, National University of the Northeast, Corrientes, Argentina}

\author{John Guillamon}
\thanks{These authors contributed equally to this work}%
\affiliation{Wave Transport in Complex Systems Lab, Department of Physics, Wesleyan University, Middletown, Connecticut 06459, USA}

\author{Krishna Joshi}
\thanks{These authors contributed equally to this work}%
\affiliation{Wave Transport in Complex Systems Lab, Department of Physics, Wesleyan University, Middletown, Connecticut 06459, USA}

\author{Tsampikos Kottos}
\email{tkottos@wesleyan.edu}
\affiliation{Wave Transport in Complex Systems Lab, Department of Physics, Wesleyan University, Middletown, Connecticut 06459, USA}

\begin{abstract}
Developing self-optimizing physical systems that learn directly from their response to external stimuli offers a promising route toward adaptive and intelligent machines. A key ingredient is the ability to evaluate optimization gradients within the physical hardware, bypassing the need for accurate numerical models. In wave systems, \textit{in-situ} adjoint methods provide this capability by exploiting reciprocity to physically realize the adjoint dynamics. However, this approach generally breaks down in nonreciprocal systems, where the adjoint field cannot be directly accessed through the original device. Here, we develop an \textit{in-situ} adjoint optimization framework that extends physical gradient computation to wave systems with asymmetric couplings through computationally low-cost similarity transformations between the forward and transposed operators. We utilize available degrees of freedom like tailored injected excitations, targeted measurements, and tunable parameters, to enable adjoint field reconstruction through forward experiments on the same physical platform. We demonstrate the versatility of the protocol through broadband transmission and isolation, coherent perfect absorption, programmable wavefront shaping, and spatially selective sensing. Our results provide a route to self-optimization in complex nonreciprocal wave systems through efficient physical gradient-based learning. 
The number of experiments required by the protocol does not increase with the number of optimization parameters, and the method remains applicable even when parts of the system are completely inaccessible.
% The protocol requires only two measurements per gradient evaluation, irrespective of the number of optimization parameters, and is applicable even when parts of the system are completely inaccessible.
\end{abstract}
\maketitle

\section{Introduction}

Developing inanimate physical systems capable of harnessing their structural complexity for self-optimization is a central challenge 
towards realizing intelligent and self-adapting machines. An emerging approach is to exploit the dynamics of physical systems as computational resources, allowing the hardware to participate in the learning process rather than relying exclusively on numerical models \cite{ZangenehNejad2021Analogue,wright2022deep,Hughes2019Wave,McMahon2023Physics,Momeni2025Training,Stern2023Learning}. Wave platforms are particularly attractive for this purpose because propagation and interference combine signals along many paths simultaneously, offering parallel processing with low latency and low energy consumption~\cite{wetzstein2020inference,shastri2021photonics}. In this context, structural complexity, often regarded as an obstacle to control, can instead provide a resource for engineering adaptive functionalities \cite{guillamon2025situ,wang2026guiding,li2026adjoint,guillamon2026nonlinear,kwon2026timedomain}.  Programmable photonic circuits, wavefront-shaping systems, and reconfigurable microwave surfaces provide different ways to control such intriguing wave dynamics~\cite{bogaerts2020programmable,mosk2012controlling,alexandropoulos2021reconfigurable}. However, the collective nature of wave scattering, together with its sensitivity to small perturbations, makes the identification of optimal configurations a high-dimensional optimization problem.

Gradient-based optimization provides a systematic approach to navigating high-dimensional parameter spaces and can offer substantial efficiency advantages over derivative-free strategies, such as genetic algorithms and particle-swarm optimization~\cite{sigmund2011usefulness,goldberg1989genetic}. However, these advantages depend critically on the cost of evaluating the gradient. Conventional finite-difference methods require perturbing each control parameter individually, resulting in a number of simulations or experiments that grows linearly with the number of tunable parameters~\cite{goldberg1989genetic,aage2017giga}. Adjoint sensitivity analysis overcomes this bottleneck by obtaining the complete gradient of a scalar objective from only one forward and one adjoint solution, independently of the number of control parameters~\cite{jensen2011topology,lalaukeraly2013adjoint,christiansen2021inverse,di2020reconfigurable}. In this framework, the forward field characterizes the device response, while the adjoint field encodes the information required to evaluate the gradient, following the same chain-rule principle underlying backpropagation in artificial neural networks~\cite{hughes2018adjoint}. This favorable scaling makes adjoint methods attractive for the optimization of complex wave systems with many degrees of freedom.

Conventionally, the forward and the adjoint fields are computed numerically, requiring a \textit{digital twin}, i.e., an accurate model of the physical system~\cite{wright2022deep,perez2020multipurpose}. Developing such a model can be particularly challenging in complex wave platforms, where even small discrepancies arising from fabrication imperfections or uncertainties in system parameters can lead to substantial deviations in the device response. These discrepancies can compromise the accuracy of the calculated gradients and, consequently, the effectiveness of the optimization. Moreover, the need for numerical simulations at each iteration introduces an additional computational burden.

A promising alternative is \textit{in-situ} adjoint optimization, in which the physical device itself is used to obtain the forward and adjoint fields. The latter typically involves the transpose of the operator governing the equations of motion, which in reciprocal systems can be physically realized using the same hardware, combined with appropriately tailored driving sources~\cite{hughes2018adjoint,pai2023experimentally, guillamon2025situ,kwon2026timedomain}. By performing these operations directly on the physical platform, this methodology eliminates the need for an accurate numerical model and reduces the computational burden associated with simulations.
Importantly, the gradient can be constructed from these measurements without requiring a complete characterization of the system's scattering response.

The reciprocity constraint excludes the applicability to nonreciprocal systems, encompassing devices designed for directional wave transport, such as isolators and circulators~\cite{caloz2018nonreciprocity,Fleury2015Nonreciprocal,Asadchy2020Nonreciprocity,Jalas2013Isolator,Kord2020Microwave,Fleury2014Acoustic,Yang2020LiDAR,Li2025Plasmonic,Yang2024NonreciprocalThermalPhotonics,Zhao2024UnidirectionalEP,Bi2011MagnetoOpticalIsolation}.
Among this category are linear systems with odd vector potentials and nonlinear systems that have an important role in the development of physical neural networks~\cite{guillamon2026nonlinear,li2026adjoint,spall2025training}.

Another important nonreciprocal class corresponds to networks with asymmetric couplings, exemplified by Hatano--Nelson systems realized in a variety of frameworks such as optical, mechanical, acoustical, and RF circuitry ~\cite{hatano1996localization,Maddi2024HatanoNelson,weidemann2020topological,brandenbourger2019nonreciprocal,helbig2020generalized,SheikhAnsari2023AsymmetricTransmission,Li2024DynamicSkin,Fleury2016Floquet,Zhang2021AcousticSkin}. In such systems, the adjoint dynamics involve the transposed operator, which reverses the directed couplings and is therefore not directly accessible by exciting the original device.  The central challenge is therefore to establish a practical protocol for obtaining the required adjoint fields directly from the original nonreciprocal platform, using experimentally accessible sources and measurements, without modifying the hardware throughout the optimization process.

Here, we introduce an \textit{in-situ} adjoint optimization framework for nonreciprocal non-Hermitian wave systems due to asymmetric couplings that exploits similarity transformations between the forward and transposed operators. As illustrated in Fig.~\ref{fig:fig1}, the adjoint field can be recovered through forward experiments on the original device, using appropriately designed sources and targeted measurements, without modifying the structural configuration. We establish the experimental accessibility conditions under which such protocols can be implemented.  For wave networks admitting a diagonal similarity transformation, we provide an explicit construction based solely on directional-coupling ratios, subject to a consistency condition in networks containing loops. Importantly, we show that diagonal transformations are the only ones that can remain fixed under arbitrary independent onsite tuning, making them particularly suitable for iterative optimization. This enables physical gradient-based optimization without requiring a complete modeling of the system, including objectives involving multiple frequencies, inputs, or perturbations. Interestingly, the protocol remains applicable even when parts of the system are inaccessible or unknown.

We demonstrate the versatility of our autonomous physical adjoint learning protocol through several optimization tasks in asymmetric wave networks. These include broadband transmission with suppressed reflection, broadband directional transport in coupled rings with limited tunable parameters, coherent perfect absorption in the presence of an unknown absorbing subsystem, output wavefront shaping, and spatially selective sensors. Finally, we demonstrate the broader applicability of our protocol in coupled optical ring resonators where the effective scattering wave-propagation equations are asymmetric. 

\begin{figure}
    \centering
    \includegraphics[width=0.5\textwidth]{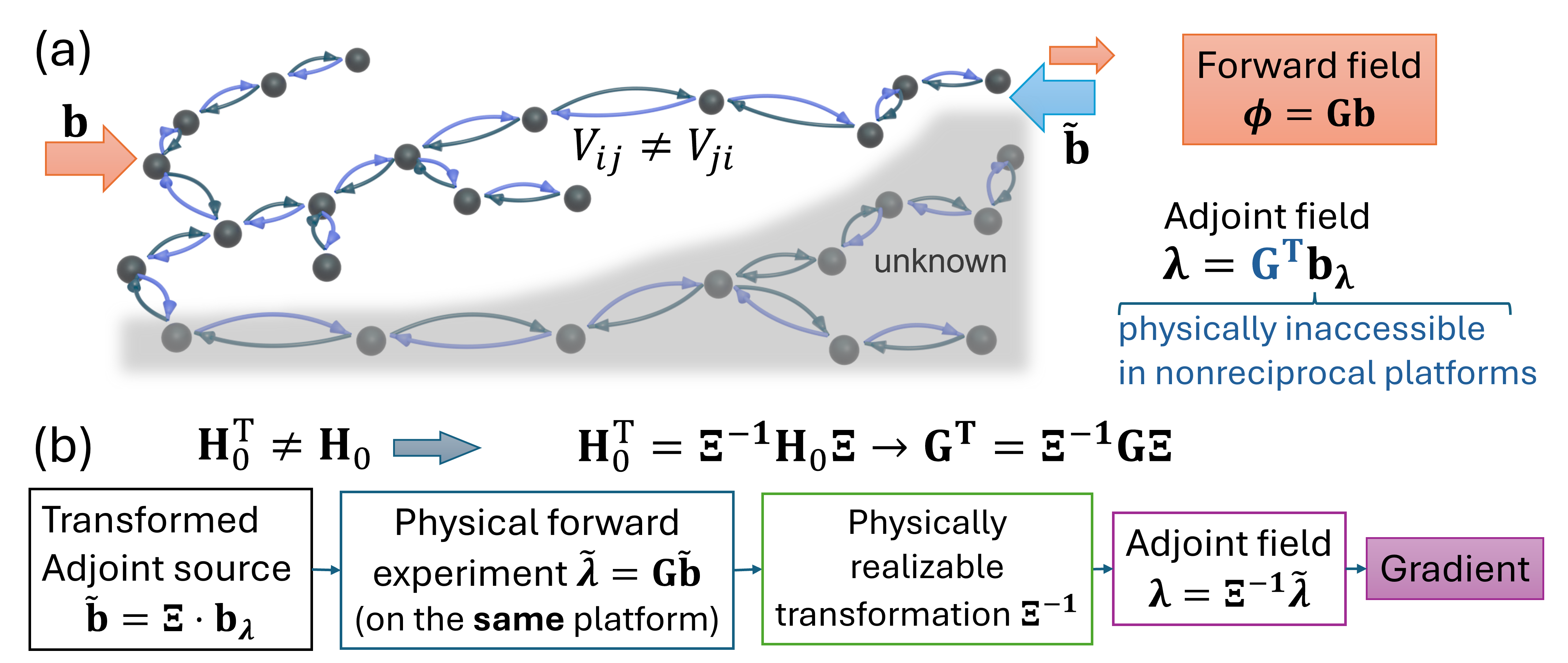}
   \caption{\textbf{In-situ adjoint optimization in asymmetric wave systems.}
(a) Schematic of a network with asymmetric directional couplings, $V_{ij}\neq V_{ji}$, and a partially unknown region (shaded). While the forward field $\boldsymbol{\phi}=\mathbf{G}\mathbf{b}$ is physically accessible, the adjoint field $\boldsymbol{\lambda}=\mathbf{G}^T\mathbf{b}_{\lambda}$ involves the transposed dynamics, which cannot generally be realized through direct excitation of the original nonreciprocal system.
(b) Proposed in-situ adjoint protocol exploiting a similarity transformation between the forward and transposed operators. The adjoint field is reconstructed from a physically realizable forward experiment on the original device, using appropriately transformed sources and measured fields, enabling direct evaluation of the gradients required for optimization.}
% (b) Proposed protocol based on a similarity transformation, $\mathbf{H}_0^T=\mathbf{\Xi}^{-1}\mathbf{H}_0\mathbf{\Xi}$, relating the forward and adjoint Green's functions. The transformed adjoint source $\widetilde{\mathbf{b}}=\mathbf{\Xi}\mathbf{b}_{\lambda}$ is injected into the original device, generating the physical response $\widetilde{\boldsymbol{\lambda}}=\mathbf{G}\widetilde{\mathbf{b}}$. The adjoint field is then reconstructed as $\boldsymbol{\lambda}=\mathbf{\Xi}^{-1}\widetilde{\boldsymbol{\lambda}}$, enabling gradient evaluation without modifying the internal dynamics of the device.}
\label{fig:fig1}
    \label{fig:fig1}
\end{figure}
%%%%%%%%%%%%%%%%%%%%%%%%%%%%%%%%%%%%%%%%%%%%%%%%%%%%%%%%%%%%%%%%%%%%%%%%
%%%%%%%%%%%%%%%%%%%%%%%%%%%%%%%%%%%%%%%%%%%%%%%%%%%%%%%%%%%%%%%%%%%%%%%%
\section{Physical adjoint optimization in asymmetric wave systems}

We consider a finite-dimensional scattering system described within a coupled-mode framework by an internal operator $\mathbf{H}_0$ and a coupling matrix $\mathbf{D}$ connecting the system to external channels.
In our optimization scheme, we allow the objective function to depend on the system response under a set of configurations $c=1,\ldots,N_c$.
A configuration represents any prescribed realization used to evaluate the objective and may correspond, for example, to a different excitation frequency, system perturbation, input state, or boundary condition.

For each configuration $c$, the steady-state field amplitudes $\boldsymbol{\phi}_c$ satisfy
\begin{equation}
\mathbf{f}_c\left(\boldsymbol{\phi}_c,\mathbf{p}\right) \equiv
\mathbf{A}_c(\mathbf{p})\boldsymbol{\phi}_c -\mathbf{b}_c =0,
\label{eq:forward_general}
\end{equation}
where, within the coupled-mode description,
\begin{equation}
\mathbf{A}_c(\mathbf{p}) =
\omega_c \mathbf{I} -\mathbf{H}_{0,c}(\mathbf{p}) +\frac{i}{2}\mathbf{D}^T\mathbf{D},
\qquad \mathbf{b}_c = i\mathbf{D}^T\mathbf{S}_c^+,
\label{eq:A_general}
\end{equation}
and the outgoing scattered field amplitudes are
\begin{equation}
\mathbf{S}_c^- = -\mathbf{S}_c^+ + \mathbf{D}\boldsymbol{\phi}_c.
\end{equation}
The forward field can equivalently be written as
\begin{equation}
\boldsymbol{\phi}_c  = \mathbf{G}_c\mathbf{b}_c, \qquad  \mathbf{G}_c=\mathbf{A}_c^{-1},
\label{eq:green_general}
\end{equation}
where $\mathbf{G}_c$ is the Green's function governing the wave propagation of an excitation at frequency $\omega_c$ under a specific configuration $c$.

\subsection{Adjoint gradients for multi-configuration objectives.}

We consider a real-valued objective of the general form
\begin{equation}
g=   g \left(  h_1,\ldots,h_{N_h}; \mathbf{p} \right),
\label{eq:general_objective}
\end{equation}
where the real-valued observables $h_n$ may depend on the fields associated with one or several configurations,
\begin{equation}
h_n
=
h_n
\left(
\left\{
\boldsymbol{\phi}_c,
\boldsymbol{\phi}_c^*
\right\}_{c=1}^{N_c};
\mathbf{p}
\right).
\end{equation}
The total derivative of the objective with respect to the real optimization parameters $\mathbf{p}$ is therefore
\begin{equation}
\dv{g}{\mathbf{p}}
=
\pdv{g}{\mathbf{p}}
+
2\Re
\left[
\sum_{c=1}^{N_c}
\pdv{g}{\boldsymbol{\phi}_c}
\pdv{\boldsymbol{\phi}_c}{\mathbf{p}}
\right],
\label{eq:total_gradient_general}
\end{equation}
where $\boldsymbol{\phi}_c$ and $\boldsymbol{\phi}_c^*$ are treated as independent variables.

Differentiating the forward constraint, Eq.~\eqref{eq:forward_general}, gives
\begin{equation}
\pdv{\boldsymbol{\phi}_c}{\mathbf{p}}
=
-\mathbf{A}_c^{-1}
\pdv{\mathbf{f}_c}{\mathbf{p}},
\end{equation}
where $\partial\mathbf{f}_c/\partial\mathbf{p}$ is evaluated at fixed $\boldsymbol{\phi}_c$.
We then introduce an adjoint field for each configuration $\boldsymbol{\lambda}_c$ through
\begin{equation}
\mathbf{A}_c^T\boldsymbol{\lambda}_c
=
\left(
\pdv{g}{\boldsymbol{\phi}_c}
\right)^T.
\label{eq:adjoint_general}
\end{equation}
Equivalently,
\begin{equation}
\boldsymbol{\lambda}_c
=
\mathbf{G}_c^T
\left(
\pdv{g}{\boldsymbol{\phi}_c}
\right)^T.
\label{eq:adjoint_green_general}
\end{equation}
The adjoint source follows directly from the dependence of the objective on the observables,
\begin{equation}
\left(
\pdv{g}{\boldsymbol{\phi}_c}
\right)^T
=
\sum_{n=1}^{N_h}
\pdv{ g}{h_n}
\left(
\pdv{h_n}{\boldsymbol{\phi}_c}
\right)^T.
\label{eq:adjoint_source_general}
\end{equation}
Thus, although the forward equations associated with different configurations are independent, their adjoint sources need not be: the coefficients $\partial g/\partial h_n$ may depend on observables obtained from all configurations.

Substitution into Eq.~\eqref{eq:total_gradient_general} yields the adjoint gradient
\begin{equation}
\dv{g}{\mathbf{p}}
=
\pdv{g}{\mathbf{p}}
-
2\Re
\left[
\sum_{c=1}^{N_c}
\boldsymbol{\lambda}_c^T
\pdv{\mathbf{f}_c}{\mathbf{p}}
\right].
\label{eq:adjoint_gradient_general}
\end{equation}
Consequently, evaluating the full gradient requires one forward and one adjoint solution per configuration, irrespective of the number of optimization parameters.
The computational cost therefore scales with the number of configurations entering the objective rather than with the dimensionality of the parameter space.
Multifrequency objectives constitute a particular case of this framework in which the configurations correspond to different excitation frequencies.

The remaining challenge for an \textit{in-situ} implementation is physical rather than computational: the adjoint field in Eq.~\eqref{eq:adjoint_green_general} is governed by the transposed Green's function $\mathbf{G}_c^T$, which in a generic nonreciprocal system does not coincide with its forward propagator.
The central question is therefore whether this adjoint evolution can be reconstructed through physically realizable operations.

%%%%%%%%%%%%%%%%%%%%%%%%%%%%%%%%%%%%%%%%%%%%%%%%%%%%%%%%%%%%%%%%%%%%%%%%%
\subsection{Physical reconstruction of adjoint fields.}

The physical realizability of the adjoint field is determined by the relation between the forward and adjoint propagators, described by $\mathbf{G}$ and $\mathbf{G}^T$, respectively (for simplicity, we drop the configuration index). 
In some systems, the relation between forward and adjoint propagation follows directly from symmetry.
For instance, in systems with a transpose-symmetric internal Hamiltonian, $\mathbf{H}_0^T=\mathbf{H}_0$, and symmetric lead self-energy $\mathbf{D}^T\mathbf{D}$, we have $\mathbf{G}^T=\mathbf{G}$.
The adjoint field can therefore be generated using the same physical system driven by the appropriate adjoint source.
More generally, in systems subject to a magnetic bias $B$, microscopic reversibility may lead to
$\mathbf{H}_{0,c}^T(B)=\mathbf{H}_{0,c}(-B)$,
and consequently
$\mathbf{G}_c^T(B)=\mathbf{G}_c(-B)$,
allowing the adjoint propagation to be realized by reversing the magnetic bias.
This relation also extends to non-Hermitian systems containing reciprocal gain or loss terms that remain invariant under transposition.
For generic asymmetric systems, however, no such direct correspondence needs to exist.

Here, we consider a broader class of asymmetric systems for which the transpose dynamics are related to the forward dynamics through a similarity transformation.
Specifically, we assume that an invertible operator $\mathbf \Xi$ exists such that
\begin{equation}
\mathbf{H}_{0}^T = \mathbf \Xi^{-1}\mathbf{H}_{0}\mathbf \Xi.
\label{eq:similarity_H}
\end{equation}
For the systems considered here, we further assume that the channel-coupling term is invariant under the same similarity transformation,
$\mathbf \Xi^{-1}\mathbf{D}^T\mathbf{D}\mathbf \Xi
= \mathbf{D}^T\mathbf{D}$.
Under this condition, the transformation of the internal operator $\mathbf{H}_0$ is directly inherited by the full scattering operator $\mathbf{A}$ and, consequently, by its Green's function $\mathbf{G}$.
Therefore, the adjoint field can be written as
\begin{equation}
\boldsymbol{\lambda} =
\mathbf \Xi^{-1}\mathbf{G} \mathbf \Xi
\left( \pdv{g}{\boldsymbol{\phi}} \right)^T.
\end{equation}
This transformation allows us to conveniently obtain the adjoint field from a physically realizable forward propagation.
\begin{equation}
\widetilde{\boldsymbol{\lambda}} = \mathbf{G}\, \widetilde{\mathbf{b}},
\end{equation}
which uses the definitions of the transformed adjoint source and adjoint field
\begin{equation}
\widetilde{\mathbf{b}} \equiv \mathbf \Xi \left( \pdv{g}{\boldsymbol{\phi}} \right)^T, 
\quad 
\widetilde{\boldsymbol{\lambda}}=\mathbf \Xi \boldsymbol{\lambda}.
\end{equation}
Thus, whenever the transformed source $\widetilde{\mathbf{b}}$ can be physically synthesized, the adjoint field can be reconstructed from a forward experiment followed by the known transformation $\mathbf \Xi^{-1}$.
This establishes a physically realizable adjoint protocol for a broad class of asymmetric systems that need not themselves be transpose symmetric.

%%%%%%%%%%%%%%%%%%%%%%%%%%%%%%%%%%%%%%%%%%%%%%%%%%%%%%%%%%%%%%%%%%%%%%%%%
\subsection{Diagonal similarity transformations: construction and applicability}
\label{sec:graph_consistency}

% The key to implementing Eq.~\eqref{eq:similarity_H} is a similarity transformation. 
Although every square matrix is similar to its transpose~\cite{taussky1959similarity}, this algebraic guarantee alone does not provide an effective \textit{in-situ} protocol. A general transformation may require access to local fields at experimentally inaccessible sites (see the Supplementary Material). In addition, efficient optimization benefits from transformations that do not require recalculation as the control parameters are tuned. Similarity transformations that act locally and can be constructed from known internal parameters are therefore particularly advantageous.

% The key to implementing Eq.~\eqref{eq:similarity_H} \textit{in-situ} is a similarity transformation that acts locally and remains known as the system is optimized. Although every square matrix is similar to its transpose~\cite{taussky1959similarity}, this algebraic guarantee alone does not provide such a protocol: a general transformation may mix accessible and inaccessible sites and require recalculation as the optimized parameters are tuned. We therefore focus on a local transformation determined by known internal coupling ratios.

A transformation that satisfies the above mentioned restrictions corresponds to a diagonal mapping $\mathbf \Xi=\operatorname{diag}(\xi_1,\ldots,\xi_N)$ that rescales each site independently with weights $\xi_n$, $n=1,\cdots, N$. For site-local probes, the mapping requires only rescaling the excitation and readout at accessible sites. Diagonal mappings are also the only similarity transformations that can remain unchanged under independent variation of every onsite term, provided that internal couplings are kept fixed (see Sec.~\ref{app:fixed_onsite_similarity}). Once constructed, the same $\mathbf \Xi$ can thus be used throughout optimization of onsite detuning, gain, and loss. 

Assuming diagonal $\mathbf \Xi$, Eq. (\ref{eq:similarity_H}) gives the recursive relation
\begin{equation}
\frac{\xi_j}{\xi_i}=\frac{(\mathbf H_0)_{ji}}{(\mathbf H_0)_{ij}},
\label{eq:edge_relation}
\end{equation}
which allows us to construct $\mathbf \Xi$ given a root $\xi_1\ne 0$. 
As a matter of fact, an invertible diagonal mapping requires nonzero coupling in both directions on every bond, i.e., $(H_0)_{ji}\ne 0$ $\forall \, (i,j)$-pairs. 

In linear chains or tree-graph networks, the recursion chain from Eq. (\ref{eq:edge_relation})  provides all elements in the diagonal of $\Xi$. However, when the recursion folds on itself, like in graph-structures having loops, these edge relations might not be consistent.
Therefore, the graph determines whether these edge relations assign a consistent set of site weights.

\textit{Chains and trees.}
Choosing one nonzero root weight fixes all remaining weights along the unique paths through the network. Any nonzero coupling assignment, therefore, admits the diagonal mapping, with no additional constraints and the weights follow directly from products of the edge ratios.

\textit{Networks with loops.}
% Different paths between the same sites must assign the same weight ratio.
For every independent cycle $\mathcal C=(i_1,\ldots,i_m,i_1)$, Eq. (\ref{eq:edge_relation}) results in a loop-condition
\begin{equation}
\prod_{k=1}^{m}
\frac{(\mathbf H_0)_{i_{k+1}i_k}}
     {(\mathbf H_0)_{i_ki_{k+1}}}=1,
\qquad i_{m+1}=i_1.
\label{eq:cycle_condition}
\end{equation}
Together with bidirectional support, this condition is necessary and sufficient for diagonal similarity. Once it holds, the weights are constructed along a spanning tree of each connected component. Importantly, loops impose additional constraints on the directional coupling ratios, but leave the onsite terms unrestricted. As the number of independent cycles increases, these constraints become more restrictive, limiting the networks that admit a diagonal similarity transformation.

For either class, fixed directional coupling ratios allow $\mathbf \Xi$ to be constructed once and used throughout onsite optimization. If the ratios are also optimized, the weights can be updated from the current known ratios using the same path products, provided looped networks retain cycle consistency. This explicit construction requires no new solution of a general matrix-similarity problem.

%%%%%%%%%%%%%%%%%%%%%%%%%%%%%%%%%%%%%%%%%%%%%%%%%%%%%%%%%%%%%%%%%%%%%%%%%
\section{Self-optimization of asymmetric wave networks}

We now apply the \textit{in-situ} adjoint formulation developed above to several representative asymmetric coupled-mode systems. We also consider objective functions that depend on multiple system configurations.

\subsection{Broadband transmission engineering}

\begin{figure}
\centering
\includegraphics[width=0.5\textwidth]{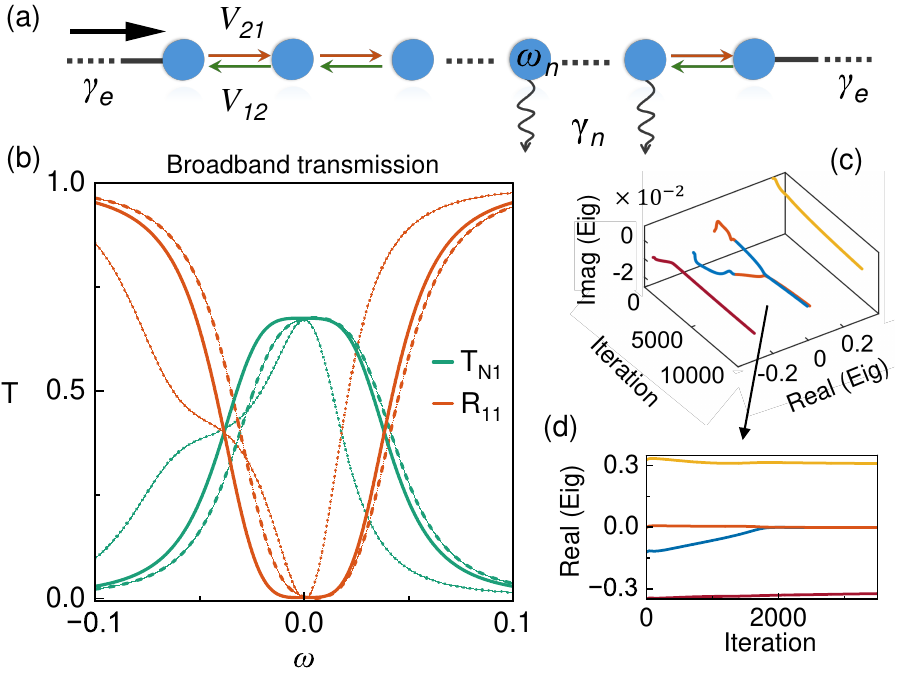}
\caption{\textbf{Broadband transmittance.}
(a) Schematic of the one-dimensional asymmetric tight-binding lattice with nonreciprocal couplings $V_{nm}\neq V_{mn}$ and external coupling $\gamma_e$. The onsite potentials $\omega_n$ are optimized in the presence of weak random loss $\gamma_n\in[0,0.001]$.
(b) Optimized transmittance (green) and reflectance (red). Dotted curves show optimization at $\omega_0=0$, while dashed and solid curves include first- and second-order spectral derivatives, respectively, producing a broader and flatter transmission window.
(c,d) Eigenvalues of the auxiliary Hamiltonian associated with the reflectionless condition. Two eigenvalues approach each other near $\omega_0$, showing that the broadband state is close to a second-order reflectionless degeneracy.
}
\label{fig:BT}
\end{figure}

We begin by considering the optimization of broadband transmittance in an asymmetric one-dimensional lattice, as shown in Fig.~\ref{fig:BT}(a). The system consists of a tight-binding chain with $N$ sites coupled at its two ends to external leads with coupling strength $\gamma_e=0.1$. The internal couplings are asymmetric, such that the forward and backward hopping amplitudes are generally different, i.e.,$V_{nm}\neq V_{mn}$, thereby enabling nonreciprocal transport and directional interference effects. Weak random onsite losses are included to account for unavoidable dissipation and losses associated with the experimental readout of the local fields. The onsite potentials $\omega_n$ are treated as optimization parameters and are updated during the optimization.

Our objective is to engineer a transmission response that is not only large at a target frequency $\omega_0$,
but also remains high and approximately flat over a finite spectral interval around it. To assess how spectral constraints affect the optimized response, we compare objective functions that progressively incorporate neighboring frequencies and penalties on the first- and second-order spectral derivatives.

%%%
We consider three objective functions involving spectral information of increasing order. In the multiconfiguration formulation of Eq.~\eqref{eq:general_objective}, each sampled frequency defines a different configuration, while the corresponding transmittances constitute the real-valued observables $h_n$. We define
$\omega_\pm=\omega_0\pm\delta\omega$, $T_0\equiv T(\omega_0)$, $T_\pm\equiv T(\omega_\pm)$.
Thus, in this application, the observables entering the objective function are simply 
$h_n\equiv T_n\equiv \frac{| [\mathbf{S}^-(\omega_n)]_{N}|^2}{| [\mathbf{S}^+(\omega_n)]_{1}|^2} = \frac{2\gamma_e}{ |S^+|^2} |[\boldsymbol{\phi}(\omega_n)]_N|^2$, being 
$|S^+|^2=\left|[\mathbf{S}^+(\omega_n)]_1\right|^2$ the incident power in the input channel coupled to site 1.

To define the cost functions, and for the sake of conciseness, we introduce the following compact notation:
two-frequency average and forward finite difference
$\langle T\rangle_+ \equiv \frac{T_0+T_+}{2}$, $\Delta_+T \equiv \frac{T_+-T_0}{\delta\omega}$;
the three-frequency average and centered finite differences
$\langle T\rangle_0 \equiv \frac{T_-+T_0+T_+}{3}$, $\Delta_0^{(1)}T \equiv \frac{T_+-T_-}{2\delta\omega}$,
$\Delta_0^{(2)}T \equiv \frac{T_+-2T_0+T_-}{\delta\omega^2}$.
Adopting such notation, the three objective functions are then 
\begin{eqnarray}
g_0 &\equiv& T_0, \nonumber\\
g_1 &\equiv& \langle T\rangle_+
-\alpha\left(\Delta_+T\right)^2, 
\label{eq:cost_functions_broadband}\\
g_2 &\equiv& \langle T\rangle_0
-\alpha\left(\Delta_0^{(1)}T\right)^2
-\beta\left(\Delta_0^{(2)}T\right)^2. \nonumber
\end{eqnarray}
The first objective, $g_0$, maximizes only the transmittance at the target frequency. The second objective, $g_1$, rewards high transmission at two neighboring frequencies while penalizing the local spectral slope. Finally, $g_2$ additionally penalizes the local spectral curvature around $\omega_0$. Because the derivative terms enter quadratically, slopes and curvatures of either sign are penalized equally. The parameters $\alpha>0$ and $\beta>0$ control the relative weights of the slope and curvature penalties. Consequently, maximizing $g_2$ favors transmission profiles that are simultaneously large, locally symmetric, and flat around $\omega_0$.

The resulting optimized transmittance (green) and reflectance (red) spectra are shown in Fig.~\ref{fig:BT}(b). When only the transmittance at $\omega_0$ is optimized through the cost function $g_0$, the system exhibits a narrow resonant transmission peak centered around the target frequency (green dotted curve). As higher-order derivative constraints are progressively incorporated into the optimization functional, the transmission profile becomes substantially broader and spectrally flatter. Suppressing the first spectral derivative (green dashed curve) already produces a significant enhancement of the transmission bandwidth, while simultaneously suppressing both first- and second-order derivatives (green solid curve) generates a broad high-transmission plateau extending over a substantially wider spectral interval around $\omega_0$. Simultaneously, the corresponding reflectance spectra (red curves) remain strongly suppressed across the same frequency range, demonstrating the emergence of broadband reflectionless transport.

To uncover the physical origin of this broadband reflection suppression, we analyze the auxiliary Hamiltonian associated with the reflectionless scattering condition~\cite{sweeney2020reflectionless,tuxbury2025synthetic}. The corresponding eigenvalue evolution is shown in Fig.~\ref{fig:BT}(c,d). Near the target frequency, two eigenvalues approach one another and form an effective second-order degeneracy close to $\omega_0$. This behavior indicates that the optimized broadband-transmission state evolves toward a reflectionless zero-mode condition associated with nearly coalescing scattering states. The proximity of these modes provides a physical mechanism for simultaneously suppressing reflection and enhancing transmission over a finite spectral interval, thereby generating robust broadband transport despite the presence of disorder and asymmetry.

\subsection{Coherent perfect absorption}

\begin{figure}
\centering
\includegraphics[width=0.5\textwidth]{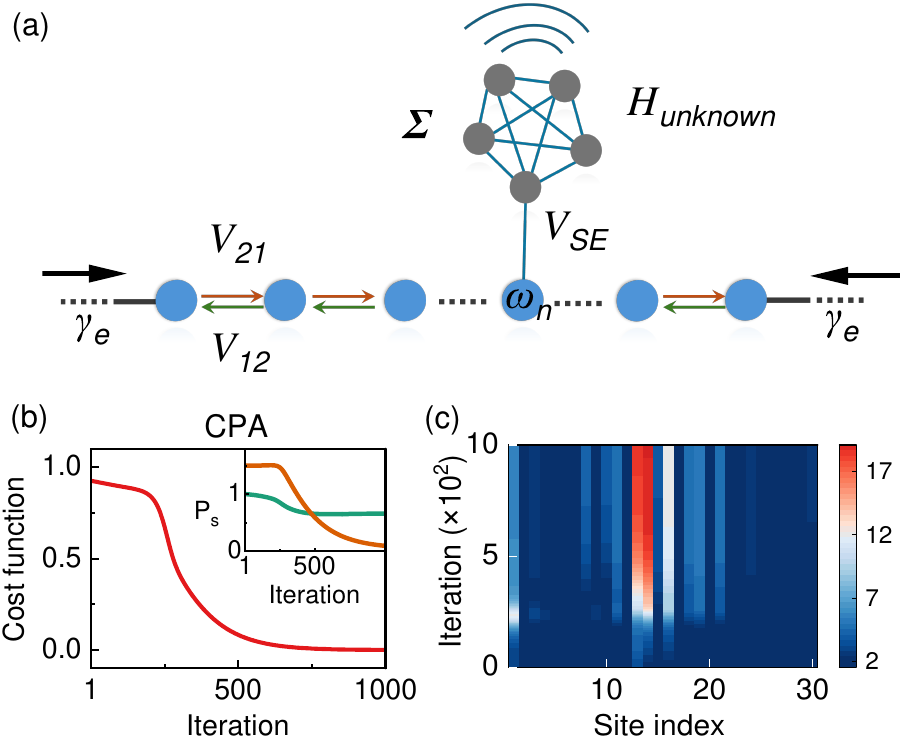}
\caption{\textbf{Coherent perfect absorption (CPA) in asymmetric lattices.}
(a) Schematic of the asymmetric lattice with an unknown absorbing subsystem (target site). The system is optimized to enhance absorption at the target site. (b) Evolution of the cost function $g=R_{11}+T_{N1}$ during optimization. Two coherent sources with amplitudes $A_s$ and phases $\theta_s$ drive the lattice. The onsite potentials $\omega_n$, together with $A_{s2}$ and $\theta_{s2}$, are optimized, while $A_{s1}=1$ and $\theta_{s1}=0$ are fixed. The inset shows the evolution of $A_{s2}$ and $\theta_{s2}$. (c) Field-intensity distribution during optimization, showing strong field buildup near the absorbing site $N_{\mathrm{abs}}=15$ and enhanced absorption in the target region.}

\label{fig:CPA}
\end{figure}

Next, we demonstrate the ability of our optimization framework to achieve coherent perfect absorption (CPA)~\cite{chong2010coherent,li2017random} in an asymmetric lattice coupled to an unknown absorbing subsystem. As illustrated in Fig.~\ref{fig:CPA}(a), the lattice is coherently excited from both directions, while the absorbing subsystem, described by the Hamiltonian $\mathbf{H}_{\mathrm{unknown}}$, is coupled to site $n=15$ through the coupling strength $V_{SE}$. While weak dissipation is present in the primary lattice, absorption occurs predominantly in the unknown subsystem. Without requiring knowledge of its internal dynamics, the optimization reshapes the onsite-potential landscape to maximize absorption in the coupled subsystem.

In this case, the lattice is driven by two coherent input channels with amplitudes $A_s$ and phases $\theta_s$. The onsite potentials $\omega_n$, together with the amplitude and phase of the second coherent source, are treated as optimization variables, while the first source is fixed at $A_{s1}=1$ and $\theta_{s1}=0$. The optimization minimizes the scattering cost function $g=R_{11}+T_{N1}$, thereby maximizing the absorption $\alpha=1-R_{11}-T_{N1}$.

Figure~\ref{fig:CPA}(b) shows the evolution of the cost function and absorption during the optimization process. The system progressively evolves toward a near-perfect absorption state characterized by vanishing reflection and transmission. The inset displays the evolution of the optimized coherent-source parameters, $A_{s2}$ and $\theta_{s2}$, illustrating how the optimization simultaneously learns the optimal interference condition required for coherent absorption. Figure~\ref{fig:CPA}(c) shows the evolution of the optimized onsite potentials $\omega_n$, revealing how the lattice landscape self-organizes to direct incoming energy toward the absorbing subsystem. Finally, Fig.~\ref{fig:CPA}(d) presents the field-intensity distribution as a function of lattice site and optimization iteration. A pronounced field buildup emerges near the absorbing site $N_{\mathrm{abs}}=15$, demonstrating strong wave localization and subsequent coherent dissipation inside the target region.

These results demonstrate that the optimized asymmetric structure can dynamically engineer interference pathways that funnel and trap incoming wave energy inside a prescribed absorbing region. The resulting CPA state arises from the combined action of coherent interference, asymmetric transport, and adaptive lattice optimization, thereby enabling near-perfect absorption through programmable wave control in complex non-Hermitian networks.

\subsection{Broadband nonreciprocal transport in coupled rings}

\begin{figure}
\centering
\includegraphics[width=0.5\textwidth]{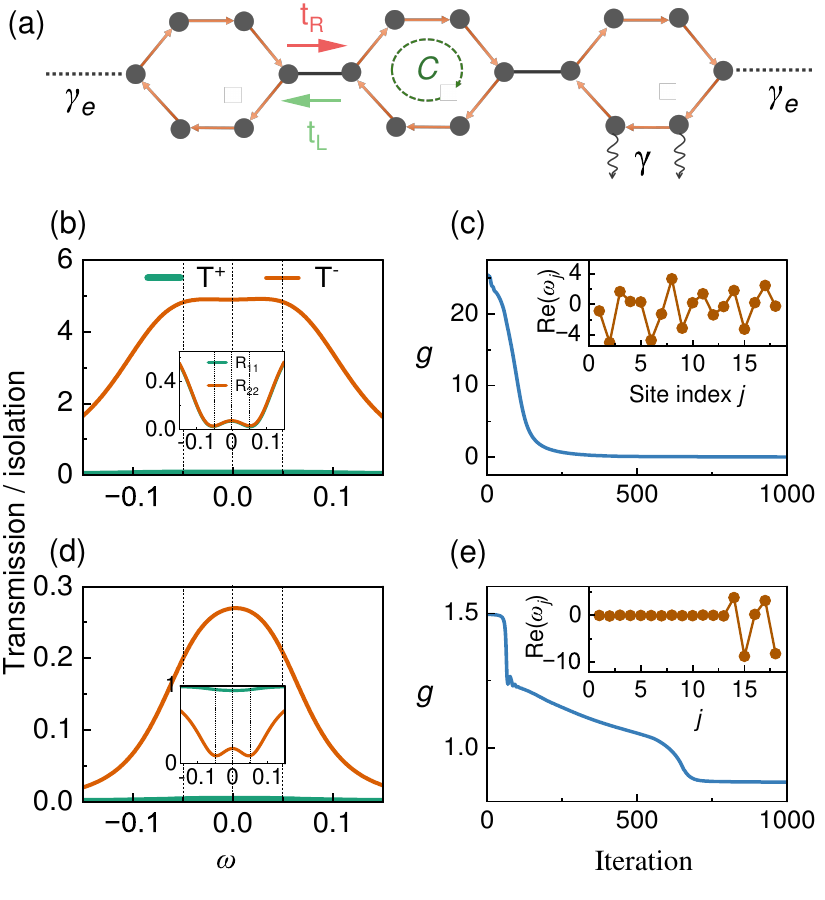}
\caption{\textbf{Broadband isolation under the cycle consistency condition.}
(a) Schematic of the coupled asymmetric-ring structure satisfying the cycle condition, with directional couplings $t_R$ and $t_L$, tunable onsite potentials $\omega_j$, uniform loss $\gamma$, and external coupling $\gamma_e$.
(b) Optimized transmission spectra showing broadband nonreciprocal transport, with $T^- \gg T^+$ around the target frequency. The inset shows the corresponding reflection spectra.
(c) Evolution of the objective function during optimization. The inset shows the optimized onsite potential profile.
(d, e) Same as in (b, c), but with control restricted to the onsite potentials of the last ring. Although the optimized transmission is lower, strong isolation ($T^- \gg T^+$) is preserved despite the reduced number of tunable parameters.}
\label{fig:isolator}
\end{figure}

We now demonstrate that the cycle consistency conditions required for diagonal similarity do not prevent successful optimization of networks containing loops. To this end, we apply the adjoint optimization protocol to engineer broadband nonreciprocal transport in a chain of coupled asymmetric rings. The system, schematically shown in Fig.~\ref{fig:isolator}(a), consists of $N_b=3$ rings, each containing $N_r=6$ resonators. Neighboring rings are connected by a single bridge link, while external ports are coupled to the left and right ends of the structure at rate $\gamma_e=0.1$.
Each ring contains asymmetric clockwise and counterclockwise hoppings, with the independent couplings initially set to $t_R=1.1$ and $t_L=0.9$, respectively. As discussed in Sec. \ref{sec:graph_consistency}, each independent cycle introduces one constraint on the hopping ratios. Accordingly, one left-going hopping in each ring is treated as a dependent parameter and chosen to satisfy the cycle condition in Eq.~\eqref{eq:cycle_condition}, thereby ensuring that the Hamiltonian obeys the diagonal similarity relation of Eq.~\eqref{eq:similarity_H}. The optimization is then performed only over the real parts of the onsite potentials, while the uniform loss $\gamma=0.02$ is kept fixed at every site.

The optimization targets broadband isolation around the central frequency $\omega_0=0$. Within the multiconfiguration framework of Eq.~\eqref{eq:general_objective}, we consider $N_c=3$ frequency configurations, $\omega_c\in\{\omega_0-\delta\omega,\omega_0,\omega_0+\delta\omega\}$, with $\delta\omega=0.05$. The corresponding directional transmittances, $T_c^{\leftarrow}$ and $T_c^{\rightarrow}$, and reflection coefficients, $R_{11,c}$ and $R_{22,c}$, constitute the real-valued observables entering the objective function. Here, $T_c^{\leftarrow}$ and $T_c^{\rightarrow}$ denote right-to-left and left-to-right transmission, respectively. We minimize
\begin{eqnarray}
g&=& \frac{1}{N_c} \sum_{c=1}^{N_c} \left[ \left(T_c^{\leftarrow}-T_{\rm tar}\right)^2 +w_{\rightarrow}\left(T_c^{\rightarrow}\right)^2 \right. \nonumber \\
&+& \left.\frac{w_R}{2}\left(R_{11,c}+R_{22,c}\right) \right].
\label{eq:loop_cost}
\end{eqnarray}
Here, $T_{\rm tar}$ is the target transmission in the preferred direction, while $w_{\rightarrow}$ and $w_R$ control the relative penalties associated with transmission in the opposite direction and reflection, respectively. Thus, minimizing $g$ drives $T_c^{\leftarrow}$ toward the desired target over the sampled frequency range while simultaneously suppressing counter-propagating transmission and back-reflection. Since the directional transmittances and reflections require excitation from both ports, each frequency configuration involves two independent forward scattering problems, one for each incidence direction.

The optimized response is shown in Fig.~\ref{fig:isolator}(b). The device develops a pronounced nonreciprocal transmission window around $\omega=0$, with $T^{\leftarrow}\approx 4.9$ and $T^{\rightarrow}\approx 0.06$ near the center of the band, closely approaching the prescribed target $T_{\rm tar}=5$ in the preferred direction. Unlike objectives based solely on the magnitude of the directional transmission contrast, Eq.~\eqref{eq:loop_cost} explicitly selects the propagation direction and drives its transmission toward a prescribed target while suppressing transport in the opposite direction. The inset of Fig.~\ref{fig:isolator}(b) shows that the reflection coefficients remain small over the optimized bandwidth, indicating that the observed isolation does not arise simply from strong back-reflection.

Figure \ref{fig:isolator}(c) shows the evolution of the objective function during optimization. The inset displays the optimized real onsite-potential profile, which develops a nonuniform spatial structure across the coupled-ring network, suggesting that the optimized response relies on collective interference rather than modifications localized to a single region. The convergence of the objective function, together with the resulting broadband nonreciprocal response, demonstrates that the cycle consistency constraints do not preclude effective optimization of asymmetric networks containing loops.

We further demonstrate that successful optimization does not require control over the entire network. To this end, we consider a more restrictive scenario in which only the onsite potentials of the last ring can be tuned. Thus, broadband transport and strong nonreciprocity are targeted by controlling only six of the eighteen onsite potentials. The results are shown in Figs.~\ref{fig:isolator}(d,e). In this case, although the maximum transmittance is reduced to $T^{\leftarrow}\approx 0.27$, a broadband transmission window is still obtained while preserving strong isolation, with $T^{\leftarrow}\gg T^{\rightarrow}\approx 0.003$. In the optimized structure, reflection is dramatically reduced only on the side for which the structure was optimized. Together, these results demonstrate that effective adjoint optimization remains possible in networks subject to cycle consistency constraints, even when control is restricted to a small subset of the onsite parameters.

\subsection{Coherent wavefront shaping in asymmetric trees}

\begin{figure*}
\centering
\includegraphics[width=1\textwidth]{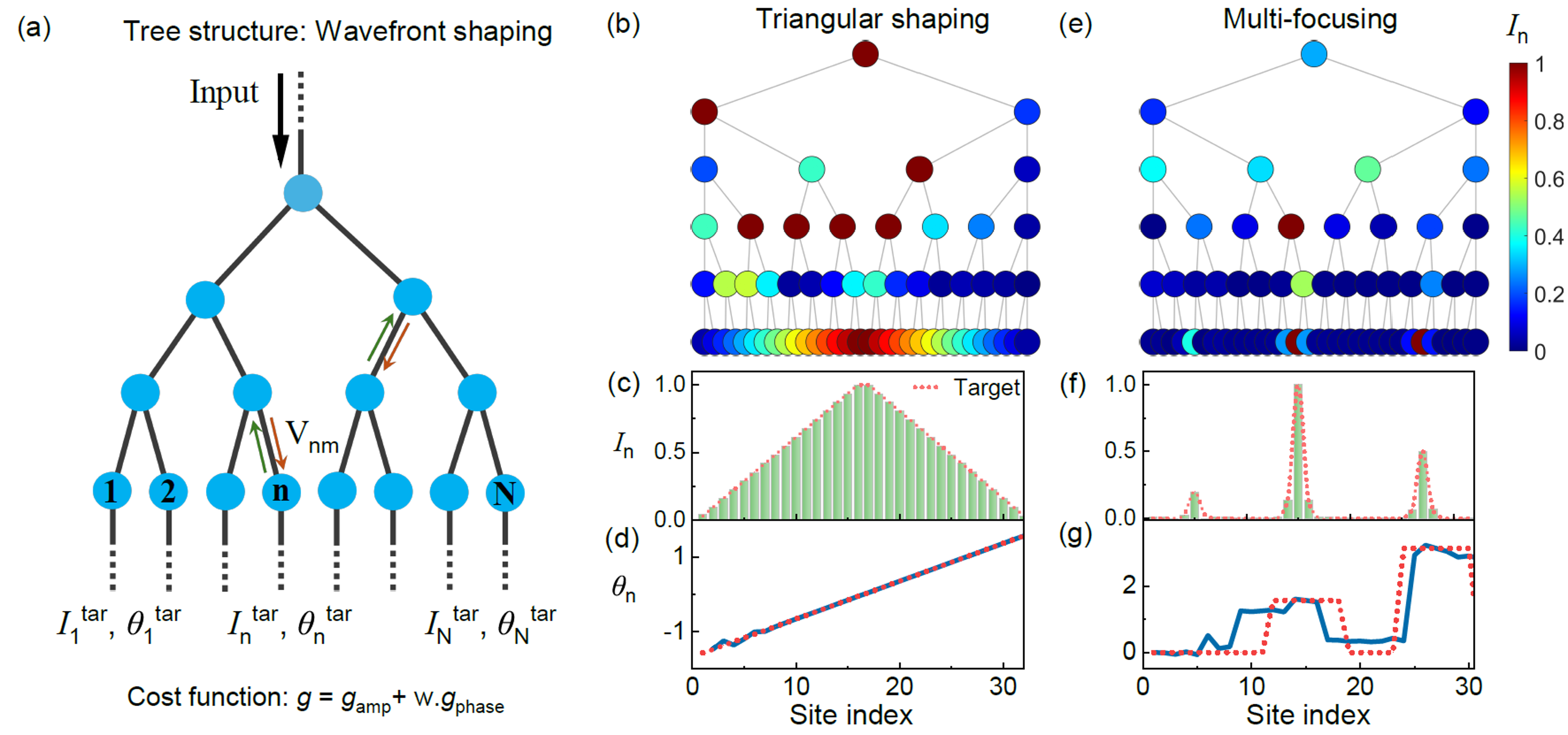}
\caption{\textbf{Programmable coherent wavefront shaping in asymmetric tree networks.}
(a) Schematic of the asymmetric tree network with tunable onsite potentials and directional couplings.
(b--d) Wavefront shaping toward a triangular target intensity profile and a linear phase varying from $-\pi/2$ to $\pi/2$: (b) optimized intensity distribution throughout the network, and comparison of the optimized and target (c) output intensity and (d) phase profiles.
(e--g) Multi-focus wavefront shaping at three target positions $n_m=(5,15,27)$ with prescribed phases $\theta_m=(0,\pi/2,\pi)$: (e) optimized intensity distribution throughout the network, and comparison of the optimized and target (f) output intensity and (g) phase profiles.}
\label{fig:wavefront}
\end{figure*}

%%%%%%%%%%%%%%%%%%%%%%%%%%%%%%%%%%%%%%%%%%%%%%%
Next, we demonstrate that the adjoint optimization framework can be extended beyond transport engineering to realize programmable coherent wavefront shaping in asymmetric networks. We consider the tree-like structure shown in Fig.~\ref{fig:wavefront}(a), where an incident wave propagates through a hierarchy of branching sites connected by asymmetric couplings, $V_{nm}\neq V_{mn}$. Because the coupling graph is a tree, no cycle consistency constraints are required for diagonal similarity, allowing the asymmetric couplings to be varied independently as long as they remain nonzero in both directions.
For demonstration purposes, we consider both the onsite potentials $\omega_n$ and the asymmetric couplings as optimization parameters, while the onsite losses are randomly assigned and kept fixed throughout the optimization. As the coupling ratios change, the similarity transformation $\mathbf{\Xi}$ is updated at each iteration directly from the instantaneous hopping amplitudes. The goal is to engineer the interference pathways throughout the network such that the field at the terminal sites reproduces a prescribed complex wavefront, enabling simultaneous control of its spatial amplitude and phase distributions. We have also verified that the protocol remains effective when control is restricted to the onsite potentials and couplings of the last two layers (results not shown)

The optimization minimizes an objective function that compares the generated output field with a prescribed target wavefront at the terminal layer, $\phi_n^{\rm tar}\equiv\sqrt{I_n^{\rm tar}}e^{i\theta_n^{\rm tar}}$, where $I_n^{\rm tar}$ and $\theta_n^{\rm tar}$ are the target intensity and phase, respectively. Specifically, we define $g=g_{\mathrm{int}}+w\,g_{\mathrm{phase}}$, where $g_{\mathrm{int}}=\sum_n\left(|\phi_n|^2-I_n^{\rm tar}\right)^2$ quantifies the intensity mismatch, while $g_{\mathrm{phase}}=\sum_n\left(\theta_n-\theta_n^{\rm tar}\right)^2$ measures the phase mismatch. Here, $\phi_n=[\boldsymbol{\phi}]_n$ is the complex field amplitude at output site $n$, $\theta_n=\arg(\phi_n)$ is its phase, and $w$ controls the relative weight of phase matching. The optimization thus drives the output field toward the prescribed intensity and phase distributions.

% The optimization is performed by minimizing an objective function that compares the generated output field with a prescribed target wavefront at the terminal layer of the network, $\phi_n^{tar}\equiv \sqrt{I_n^{tar}} e^{i\theta_n^{tar}}$, where $I_n^{tar}$ and $\theta_n^{tar}=\arg(\phi_n^{tar})$ are the corresponding target intensity and phase. Specifically, we define $g=g_{\mathrm{int}}+w\,g_{\mathrm{phase}}$, where $g_{\mathrm{int}}=\sum_n\left(|\phi_n|^2-I_n^{\rm tar}\right)^2$ quantifies the deviation of the output intensity from the target profile, while $g_{\mathrm{phase}}=\sum_n\left(\theta_n-\theta_n^{\rm tar}\right)^2$ measures the mismatch between the generated and target phase profiles. Here, $\phi_n=[\boldsymbol{\phi}]_n$ is the complex field amplitude at output site $n$, $\theta_n=\arg(\phi_n)$ is its phase, and $w$ controls the relative weight of phase matching with respect to intensity shaping. The optimization therefore drives the output field toward the prescribed intensity and phase distributions.

Figures~\ref{fig:wavefront}(b--d) and (e--g) show two representative examples of coherent wavefront shaping obtained with $w=0.25$. In the first example, the optimization targets a triangular intensity distribution together with a linear phase profile across the output layer. Specifically, the target intensity is defined as
$I_n^{\rm tar}=1-|n-n_c|/(N/2)$,
with $n_c=15$ denoting the center of the output layer, while the target phase $\theta_n^{\rm tar}$ varies linearly over the interval $[-\pi/2,\pi/2]$. Such phase-gradient profiles are commonly employed in phased-array beam steering and directional wavefront control. Figure~\ref{fig:wavefront}(b) shows the optimized two-dimensional intensity distribution throughout the tree network, illustrating how the optimization redistributes the field to synthesize the prescribed output wavefront. Figures~\ref{fig:wavefront}(c,d) compare the optimized output intensity and phase profiles with their corresponding targets, showing close agreement across the output channels.

In the second example, the optimization realizes coherent multi-focus wavefront shaping, corresponding to the programmable generation of multiple spatially separated coherent hotspots with independently controlled amplitudes and phases. The target field is constructed as
\begin{equation*}
\phi_n^{\rm tar}
= \sqrt{I_n^{tar}} e^{i\theta_n^{tar}} =
\sum_{m=1}^{M}
A_m
e^{ -\frac{(n-n_m)^2}{2\sigma_m^2}} e^{i\theta_m},
\end{equation*}
where $n_m$ defines the location of the $m$-th focal spot, $A_m$ its amplitude, $\theta_m$ its coherent phase, and $\sigma_m = 0.5$ the spatial width of the focus. Figures~\ref{fig:wavefront}(e--g) show the resulting optimized state for three prescribed focal positions located at $n=(5,15,27)$ with target phases $\theta_m=(0,\pi/2,\pi)$. 
The optimized wavefront simultaneously concentrates energy into the selected output channels while closely reproducing the prescribed phase relations. The remaining phase deviations occur predominantly at sites where the field intensity is very small and the phase is consequently poorly defined. The successful shaping of both amplitude and phase profiles highlights the versatility of the adjoint protocol, which remains applicable even when the directional couplings are optimized and the similarity transformation must be updated at each iteration.

\subsection{Spatially selective sensing in a Hatano-Nelson chain \label{sec:spatial_sensor_maintext}}

\begin{figure}[t]
\centering
\includegraphics[width=0.5\textwidth]{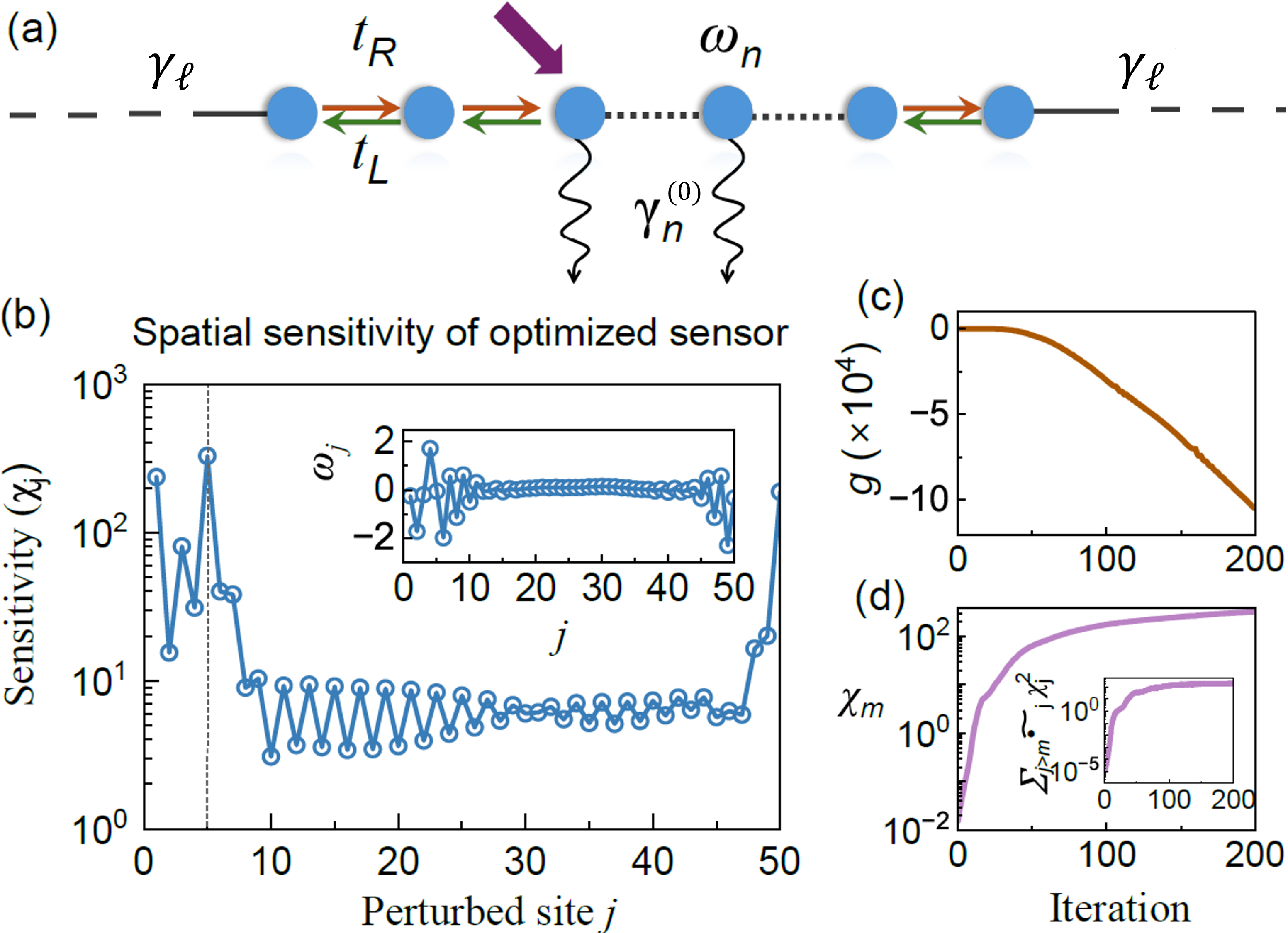}
\caption{\textbf{Spatially selective sensing in a Hatano--Nelson chain.}
(a) Schematic of the sensing configuration, with a local perturbation applied at the target site $m$ (violet arrow).
(b) Optimized spatial sensitivity $\chi_j$ as a function of the perturbed site $j$, showing a strong response at the target site marked by the dashed line. The inset shows the optimized onsite potential profile $\omega_j$.
(c) Evolution of the objective function during optimization.
(d) Target sensitivity $\chi_m$ as a function of iteration. The inset shows the weighted off-target contribution, $\sum_j \widetilde{w}_j\chi_j^2$.}
\label{fig:HN}
\end{figure}

We next demonstrate the ability of our multiconfiguration adjoint framework to engineer spatially selective sensitivity, enhancing the response to perturbations at a target location while suppressing it elsewhere. To this end, we consider a Hatano--Nelson (HN) chain of $N$ sites coupled to leads at both ends [Fig.~\ref{fig:HN}(a)]. The effective Hamiltonian $\mathbf{H}_{\rm eff}$ features asymmetric nearest-neighbor couplings $(\mathbf{H}_{\rm eff})_{j+1,j}=t_R$ and $(\mathbf{H}_{\rm eff})_{j,j+1}=t_L$, along with complex onsite potentials $\tilde{\omega}_j=\omega_j-i\gamma_j$. The imaginary parts account for intrinsic losses $\gamma_j^{(0)}$ and coupling to the leads at the boundary sites $j=1,N$, such that $\gamma_j=\gamma_j^{(0)}+\gamma_\ell(\delta_{j,1}+\delta_{j,N})$.

% Our goal is to engineer a sensor that is highly responsive to a perturbation at a target site $m$, while remaining comparatively insensitive to perturbations elsewhere. To probe the local sensitivity at site $j$, we consider two configurations in which the onsite potential is perturbed as ${\tilde \omega}_j\rightarrow {\tilde \omega}_j\pm\Delta p(1-i)$. 
% Thus, each spatial location is associated with two distinct configurations in the multiconfiguration framework introduced above.
% The system is excited from the left, with the input coupled to site $1$ and the transmitted signal read at site $N$. 
% For each configuration, the corresponding transmittances are denoted by
% $T_j^{(\pm)}\equiv T_{N1}\big|_{\omega_j\rightarrow\omega_j\pm\Delta p(1-i)}$, where 
% $T_{N1}=4\gamma_\ell^2|(\mathbf{G})_{N1}|^2$,
% and the Green's function $\mathbf{G}$ is evaluated at the corresponding on-site-potential configuration $\omega_j\rightarrow\omega_j\pm\Delta p(1-i)$.

To probe the local sensitivity at site $j$, we consider two configurations in which the complex onsite potential is perturbed as $\tilde{\omega}_j\rightarrow\tilde{\omega}_j\pm\Delta p(1-i)$. Thus, each spatial location is associated with two distinct configurations within the multiconfiguration framework introduced above. The system is excited from the left, with the input coupled to site $1$ and the transmitted signal measured at site $N$. The corresponding transmittances are $T_j^{(\pm)}\equiv T_{N1}\big|_{\tilde{\omega}_j\rightarrow\tilde{\omega}_j\pm\Delta p(1-i)}$, where $T_{N1}=4\gamma_\ell^2|(\mathbf{G})_{N1}|^2$.

For a perturbation applied at site $j$, we define the finite-difference sensitivity as
\begin{equation}
\chi_j= \left| \frac{T_j^{(+)}-T_j^{(-)}}{2\Delta p} \right|.
\label{eq:chi_main}
\end{equation}
The optimization seeks to maximize the sensitivity at the target site $m$ while suppressing the response outside its immediate neighborhood. To this end, we minimize
\begin{equation}
g=-\chi_m^2+\sum_{j=1}^{N}\widetilde w_j\chi_j^2,
\label{eq:cost_main}
\end{equation}
where the normalized weights exclude a five-site window centered at the target, $\widetilde w_j=0$ for $j=m-2,\ldots,m+2$ and $\widetilde w_j=1/(N-5)$ otherwise.
The onsite parameters $\omega_j$ and $\gamma_j^{(0)}$ are optimized within the bounds
$|\omega_j|\le W_{\max}=3$ and $\gamma_{\min}=0.01\le\gamma_j^{(0)}\le\gamma_{\max}=0.2$.

The optimized configuration exhibits a pronounced sensitivity peak at the target site [Fig.~\ref{fig:HN}(b)], while the response remains comparatively low over most of the penalized region. However, sizeable sensitivities persist near the lead-coupled boundaries, even when explicitly penalized during optimization. This behavior is consistent with the strong boundary sensitivity characteristic of Hatano--Nelson chains. The inset of Fig.~\ref{fig:HN}(b) shows that the optimized onsite-potential profile develops its strongest spatial variations near the boundaries and the target site, suggesting that these regions play an important role in shaping the sensitivity response. The optimized losses converge to their lower bound, $\gamma_j=\gamma_{\min}=0.01$, for all sites. Figures~\ref{fig:HN}(c,d) show the evolution of the objective function, target sensitivity, and weighted off-target contribution during optimization. 
Beyond optimizing the scattering response itself, this example highlights the ability of the multiconfiguration adjoint framework to tailor how a wave system responds to spatially distributed perturbations, enabling selective enhancement and suppression of local sensitivities.

\section{Extending the adjoint optimization to asymmetric scattering operators}

\begin{figure*}
\centering
\includegraphics[width=1\textwidth]{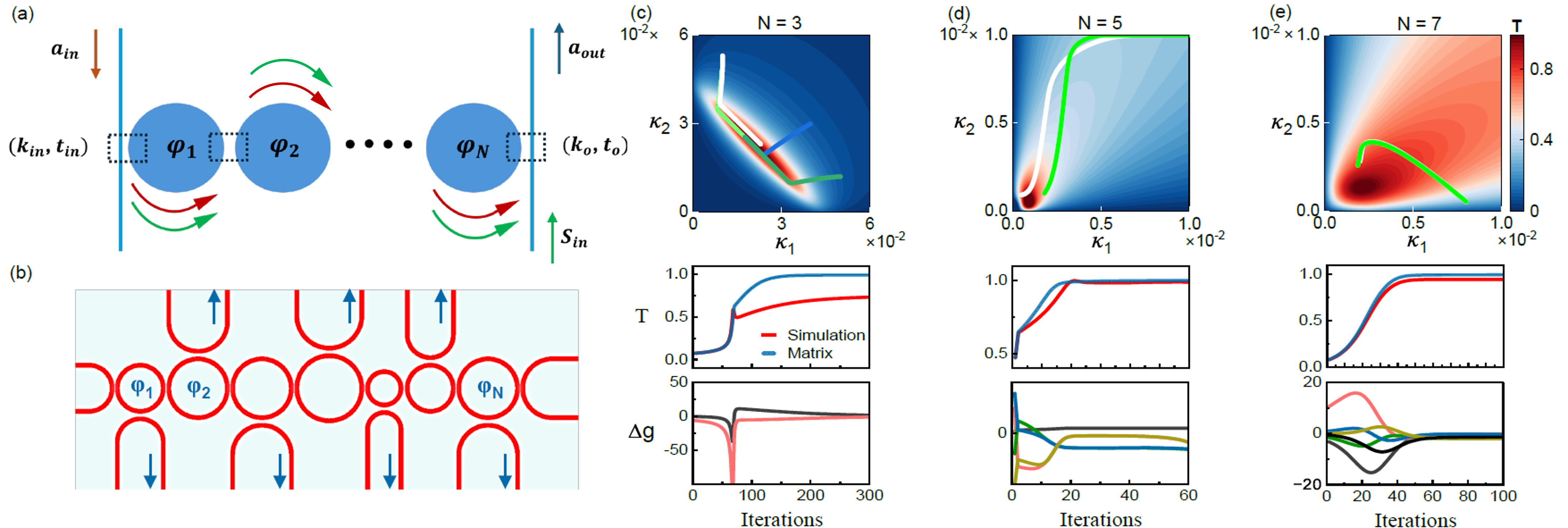}
\caption{\textbf{In-situ adjoint optimization in a chain of coupled optical ring resonators:}
(a) Schematic of an $N$-ring resonator network coupled to input and output waveguides. The intracavity field amplitudes $\{\phi_n\}$ interact via nearest-neighbor couplings $(\kappa_n, t_n)$, with input excitation $a_{\mathrm{in}}$ and output transmission $a_{\mathrm{out}}$. A backward injection $s_{\mathrm{in}}$ is used to construct the adjoint field required for gradient evaluation.
(b) Corresponding implementation in Lumerical, illustrating the cascaded ring architecture with directional couplers and bus waveguides.
(c--e) Gradient-based optimization for increasing network sizes ($N=3,5,7$). Top panels show the transmission landscape $T(\kappa_1,\kappa_2)$, spanning the full parameter space for $N=3$ and a two-dimensional cross section for $N=5,7$. Colored curves indicate optimization trajectories obtained using the proposed method, while white curves represent the full matrix reference. Middle panels compare the evolution of transmission $T$ during optimization using the proposed method (red) and the matrix reference (blue). Bottom panels show the evolution of the gradient discrepancies $\Delta g$. The results demonstrate that the proposed method successfully enhances transmission across different network sizes, achieving performance comparable to the full matrix reference.
}
\label{fig:ring_adjoint}
\end{figure*}

We now explore the broader applicability of our adjoint optimization protocol to wave systems described by asymmetric scattering operators (e.g., transfer matrices). As a representative photonic platform, we consider a chain of $N$ coupled optical ring resonators connected to input and output waveguides [Fig.~\ref{fig:ring_adjoint}(a)]. Light injected from the left with amplitude $a_{\mathrm{in}}$ excites the resonator network and is collected at the output with amplitude $a_{\mathrm{out}}$. Adjacent rings are connected through directional couplers characterized by through- and cross-coupling amplitudes $t_n$ and $\kappa_n$, respectively. We describe the internal state of the network through the intracavity amplitudes $\{\phi_n\}$. An additional excitation $s_{\mathrm{in}}$ can be injected from the opposite port to reconstruct the adjoint response required for gradient evaluation. The corresponding Lumerical implementation [Fig.~\ref{fig:ring_adjoint}(b)] reproduces this cascaded architecture, providing a realistic photonic setting for investigating the proposed optimization protocol.

The intracavity amplitudes $\{\phi_n\}$ are determined by the coherent
multiple scattering of light throughout the resonator network. Propagation
along half of ring $n$ is described by the complex factor
$P_n=e^{i\theta_n/2}$, which accounts for both phase accumulation and
attenuation or amplification. Light injected into the network can undergo
multiple round trips within each resonator and successive coupling events
between adjacent rings, giving rise to interference between different
propagation paths. After eliminating the complementary intracavity fields,
the steady-state scattering problem results
\begin{equation}
\mathbf{M}\boldsymbol{\phi}
=
\mathbf{D}\mathbf{s}^{+},
\qquad
\mathbf{G}=\mathbf{M}^{-1}, \label{eq:scattering_ring_resonators}
\end{equation}
where $\boldsymbol{\phi}=(\phi_1,\ldots,\phi_N)^T$ and
$\mathbf{s}^{+}=(a_{\rm in},s_{\rm in})^T$ collect the intracavity
and externally injected fields, respectively.

Although the underlying photonic circuit is reciprocal, the reduced scattering problem in Eq.~\eqref{eq:scattering_ring_resonators} is governed by an asymmetric matrix $\mathbf M$. As we show below, in the weak-coupling regime this matrix approximately admits the diagonal similarity transformation underlying our adjoint protocol.
Specifically, for the cascaded ring network, the exact reduced matrix $\mathbf{M}$ has
an upper-Hessenberg structure (see Supplementary Material for the full
derivation). Its diagonal elements,
\begin{equation}
M_{nn}=t_{n-1}t_nP_n^2-1,
\qquad n=1,\ldots,N,
\end{equation}
describe the round-trip recirculation within each ring, with
$t_0\equiv t_{\mathrm{in}}$ and $t_N\equiv t_{\mathrm{out}}$.
The lower nearest-neighbor elements are
$M_{n+1,n}=i\kappa_nP_n$, whereas the corresponding upper
nearest-neighbor elements are
$M_{n,n+1}=i\,t_{n-1}\kappa_n t_{n+1}P_nP_{n+1}^2$.
More generally, the higher-order upper-triangular elements arise from
successive scattering events through the chain and, for $m>n$, we have
\begin{equation}
M_{nm}
=
i^{\,m-n}t_{n-1}t_mP_m^2
\prod_{\ell=n}^{m-1}(\kappa_\ell P_\ell).
\end{equation}
Thus, $M_{n,n+r}$ contains $r$ cross-coupling amplitudes and is of order
$O(\kappa^r)$. These effective longer-range terms do not represent
direct couplings between nonadjacent resonators, but arise from multiple
successive scattering events in the underlying nearest-neighbor network.

For a given excitation, the intracavity field is obtained as
$\boldsymbol{\phi}=\mathbf{G}\mathbf{D}\mathbf{s}^{+}$. We focus on
forward excitation, $s_{\mathrm{in}}=0$, and take the transmittance
$T=\frac{|a_{\mathrm{out}}|^2}{|a_{\mathrm{in}}|^2}$
as the optimization objective. Transmission provides a direct measure of
how efficiently waves injected at one end of the network are transferred
to the opposite port and therefore constitutes a natural proxy for the
ability of the system to transport energy or information.
Optimizing $T$ therefore amounts to controlling the interference of the different
propagation paths by tuning the inter-resonator coupling coefficients
$\{\kappa_n\}$.

For the asymmetric ring network, the adjoint response involves
$\mathbf{G}^{T}$, which is not directly accessible from backward
propagation in the same structure. As shown below, in the weak-coupling
regime the structure of $\mathbf{M}$ enables $\mathbf{G}^{T}$ to be
reconstructed from the physically accessible response $\mathbf{G}$.

In the weak-coupling regime, $\kappa_n\ll1$ and
$t_n=\sqrt{1-\kappa_n^2}\simeq1$. Neglecting terms of order
$\kappa^2$ and higher reduces $\mathbf{M}$ to a tridiagonal matrix, with
$M_{n+1,n}=i\kappa_nP_n$ and
$M_{n,n+1}=i\kappa_nP_nP_{n+1}^2$. Their ratio,
$\alpha_{n+1,n}\equiv M_{n+1,n}/M_{n,n+1}=P_{n+1}^{-2}$,
defines a local asymmetry that accumulates multiplicatively along the
chain. Consequently, the tridiagonal approximation admits the diagonal similarity transformation introduced above, allowing the adjoint response to be reconstructed from forward measurements through a known site-dependent rescaling. Although this relation is not exact for the full reduced operator, the resulting approximate adjoint gradients can still provide effective optimization directions, as demonstrated below.

To validate the proposed optimization method, we implement the full ring-resonator network in Lumerical. The coupling coefficients $\kappa_n$ are treated as tunable parameters, while the cavity lengths and losses are fixed ($L_n \sim 30~\mu$m and $\gamma_n=0.1~\mathrm{dB/m}$), and the operating wavelength is set to $1.5~\mu$m. A weakly coupled probe ($\kappa\approx 0.1$) is attached to each ring to monitor the intracavity amplitudes $\{\phi_n\}$ without significantly perturbing the system.

We evaluate the optimization method for network sizes $N=3,5,7$, as shown in Fig.~\ref{fig:ring_adjoint}(c--e). The top panels display the transmission landscape $T(\kappa_1,\kappa_2)$, together with the optimization trajectories obtained using the proposed method (colored curves) and the full matrix model (white curves). For $N=3$, the landscape spans the complete optimization parameter space, whereas for $N=5$ and $N=7$ it represents a two-dimensional cross section. In all cases, the proposed method drives the system toward regions of high transmission, demonstrating that the estimated gradients provide effective directions for optimization even as the dimensionality of the parameter space increases.

The middle panels show the evolution of the transmission during optimization, comparing the proposed method with the full matrix reference. For all three network sizes, the transmission increases and eventually saturates, confirming successful optimization. Although differences in the converged transmission are observed, particularly for $N=3$, the proposed method consistently achieves substantial transmission enhancement. The bottom panels show the evolution of the gradient discrepancies $\Delta g$ between the approximate and full-matrix approaches. Despite these discrepancies, the proposed method consistently improves transmission, confirming that the approximate gradients provide effective optimization directions.

The successful optimization of these photonic networks highlights the broader applicability of our methodology to systems described by asymmetric reduced operators.

\section{Conclusions}

We have developed an \textit{in-situ} adjoint optimization protocol for wave networks of asymmetrically coupled resonators that enables the reconstruction of adjoint fields through appropriately designed forward experiments performed on the same physical platform. The reconstruction relies on a computationally low-cost diagonal similarity transformation constructed directly from the ratios of directional couplings, allowing the required adjoint sources and fields to be obtained through local rescaling. We have established the conditions under which this protocol is \textit{in-situ} viable for chains, trees, networks with loops and/or unknown portions of the total system. We have further distinguished between two groups of training parameters: (a) Resonator detunings, gain, and loss where the transformation remains unchanged; (b) The optimization involves variations in the directional couplings where the similarity transformation has to be updated at every iteration step. We have illustrated the versatility of the physical adjoint learning protocol through several optimization tasks, including broadband transmission and isolation, coherent perfect absorption involving an unknown (or inaccessible) absorbing subsystem, coherent wavefront shaping, and spatially selective hypersensitive sensing. 

Our results provide a practical framework for extending \textit{in-situ} adjoint optimization to a class of nonreciprocal wave-scattering systems, provided that the required transformations are compatible with the available sources and measurements. More broadly, they suggest opportunities for adaptive control and self-optimization of complex inanimate wave devices without requiring complete knowledge of their internal dynamics.

{\it Note added:} While this manuscript was under final stages of preparation, an independent study developing an abstract framework for physical adjoint optimization, including certain classes of nonreciprocal systems, was made available on arXiv~\cite{bosch2026unifying}.

\begin{acknowledgements}
All authors acknowledge partial support from Simons Foundation  SFI-MPS-EWP-00008530-08. 
L.J.F.-A. acknowledges the hospitality of Wesleyan University, where part of this work was
developed, and partial support from CONICET and former MINCyT grant number CONVE-2023-10189190-FFFLASH.
J.G., K.J. and T.K. acknowledge partial support from MURI ONR-N000142412548. 
\end{acknowledgements}

% \section{Conflicts of Interest}

% The authors declare no conflicts of interest.

% \section{Data Availability Statement}

% The data that support the findings of this study are available from the corresponding author upon reasonable request. 

% \nocite{*}

% \bibliography{apssamp}% Produces the bibliography via BibTeX.
\bibliographystyle{unsrt} 
\bibliography{apssamp}

% Begin the Supplementary Information on a fresh page.
\clearpage
\onecolumngrid
%%%%%%%%%% Merge with supplemental materials %%%%%%%%%%
%%%%%%%%%% Prefix a "S" to all equations, figures, tables and reset the counter %%%%%%%%%%
\setcounter{equation}{0}
\setcounter{figure}{0}
\setcounter{table}{0}
\setcounter{page}{1}
% \makeatletter
\renewcommand{\theequation}{S\arabic{equation}}
\renewcommand{\thefigure}{S\arabic{figure}}
\renewcommand{\thetable}{S\arabic{table}}
% Print an S prefix in citations and bibliography labels in this section.
\renewcommand{\citenumfont}[1]{S#1}
\renewcommand{\bibnumfmt}[1]{[S#1]}
%%%%%%%%%% Prefix a "S" to all equations, figures, tables and reset the counter %%%%%%%%%%

\section*{Supplementary Information}

% \subsection{Single-channel transport objective and adjoint excitation}

% To make the formalism concrete, we consider optimizing transmission into a target output channel $n$, with objective $g \equiv T = |S_n^-|^2/|S_1^+|^2$. For single-port excitation, this reduces to $g \propto \gamma_n |\phi_n|^2$, where $\gamma_n$ is the coupling rate to channel $n$. The corresponding derivative $\partial g / \partial \phi$ is therefore localized at the target site, implying that the adjoint source is injected at the same channel whose transmission is being optimized.

% Physically, the adjoint field corresponds to a scattering experiment driven from the output port. Specifically, an excitation satisfying $D^T \ket{\widetilde{S}^+} \propto \hat e_n$ generates an adjoint field $\lambda = i \widetilde{G} D^T \ket{\widetilde{S}^+}$, where $\widetilde{G} = (\omega I - H_0^T + \frac{i}{2}D^T D)^{-1}$. Thus, the adjoint field is governed by the transpose Green’s function and corresponds to the system’s response to a localized injection at the target channel. The central challenge is therefore to relate $\widetilde{G}$ to a physically realizable forward response.

\subsection{Adjoint sources for spectrally constrained objectives}

In this section we provide details on the adjoint sources when the objective functions depend on multiple experiments performed at different frequencies. In this case, each sampled frequency defines an independent forward problem and contributes a corresponding adjoint source. We consider the first- and second-order spectral objectives introduced in the main text as representative examples.

\subsubsection{Objective involving a first-order spectral derivative.}

We first consider
\begin{equation}
g_1=
\frac{T_0+T_+}{2}
-\alpha\left(\frac{T_+-T_0}{\delta\omega}\right)^2.
\end{equation}
Since $T_0$ and $T_+$ depend on fields evaluated at different frequencies, the objective generates one adjoint source at each frequency. Applying the chain rule gives
\begin{align}
\left(\frac{\partial g_1}{\partial\boldsymbol{\phi}_{0(+)}}\right)^T
&=
\left[
\frac{1}{2} \pm \frac{2\alpha(T_+-T_0)}{\delta\omega^2}
\right]
\left(\frac{\partial T_{0(+)}}{\partial\boldsymbol{\phi}_{0(+)}}\right)^T,
\end{align}
where the upper (lower) sign corresponds to $\omega_0$ ($\omega_0+\delta\omega$).

For transmission measured at the output port $N$,
$T_n\equiv 
\frac{| [\mathbf{S}^-(\omega_n)]_{N}|^2}
{| [\mathbf{S}^+(\omega_n)]_{1}|^2} = \frac{2\gamma_e}{ |S^+|^2} |[\boldsymbol{\phi}(\omega_n)]_N|^2$, 
with $|S^+|^2=| [\mathbf{S}^+(\omega_n)]_{1}|^2$ the same for all frequencies. 
Then,
\begin{equation}
\left(\frac{\partial T_k}{\partial\boldsymbol{\phi}_k}\right)^T
=
\frac{2\gamma_e [\phi^*(\omega_k)]_N}{|S^+|^2}
\mathbf e_N, \label{eqS:partial_T_partial_phi}
\end{equation}
with $k\in\{0,+\}$. Therefore, the adjoint sources can be constructed directly from the complex transmitted amplitudes measured in the two forward experiments performed at frequencies $\omega_0$ and $\omega_0+\delta\omega$.

The corresponding adjoint responses can then be reconstructed using the similarity relation $\mathbf G_k^T=\mathbf \Xi^{-1}\mathbf G_k\mathbf \Xi$. Since the amplitudes of the adjoint sources depend on $T_+-T_0$, they must be recalculated after each pair of forward measurements.

\subsubsection{Objective involving second-order spectral derivatives.}
We now consider
\begin{equation}
g_2 =
\frac{T_-+T_0+T_+}{3}
-\alpha\left(\frac{T_+-T_-}{2\delta\omega}\right)^2
-\beta\left(\frac{T_+-2T_0+T_-}{\delta\omega^2}\right)^2 .
\end{equation}
In this case, the three transmittances depend on fields evaluated at $\omega_0-\delta\omega$, $\omega_0$, and $\omega_0+\delta\omega$, respectively. Applying the chain rule independently at each frequency gives the three adjoint sources,
\begin{align}
\left(\frac{\partial g_2}{\partial\boldsymbol{\phi}_-}\right)^T
&=
\left[
\frac{1}{3}
+\frac{\alpha(T_+-T_-)}{2\delta\omega^2}
-\frac{2\beta(T_+-2T_0+T_-)}{\delta\omega^4}
\right]
\left(\frac{\partial T_-}{\partial\boldsymbol{\phi}_-}\right)^T, \nonumber
\\
\left(\frac{\partial g_2}{\partial\boldsymbol{\phi}_0}\right)^T
&=
\left[
\frac{1}{3}
+\frac{4\beta(T_+-2T_0+T_-)}{\delta\omega^4}
\right]
\left(\frac{\partial T_0}{\partial\boldsymbol{\phi}_0}\right)^T,
\\
\left(\frac{\partial g_2}{\partial\boldsymbol{\phi}_+}\right)^T
&=
\left[
\frac{1}{3}
-\frac{\alpha(T_+-T_-)}{2\delta\omega^2}
-\frac{2\beta(T_+-2T_0+T_-)}{\delta\omega^4}
\right]
\left(\frac{\partial T_+}{\partial\boldsymbol{\phi}_+}\right)^T. \nonumber
\end{align}
As in the first-order case, Eq.~(\ref{eqS:partial_T_partial_phi}) applies at each frequency, now with $k\in\{-,0,+\}$, so that all quantities required to construct the adjoint sources are obtained from the three forward scattering measurements.

The first-derivative penalty produces corrections of opposite sign at the two side frequencies and does not affect the source at $\omega_0$, whereas the second-derivative penalty modifies all three sources. Each optimization step therefore requires forward measurements at the three sampled frequencies, followed by the corresponding forward realizations of the three adjoint problems using the similarity transformation described above. No direct measurement of either spectral derivative is required, since both penalties are constructed from the measured values $T_-$, $T_0$, and $T_+$. The adjoint sources are recalculated after each set of forward measurements.

\subsection{Scattering formulation for a chain of coupled ring resonators}

We derive here the scattering description underlying the effective coupled-mode model used in the main text. We first consider three coupled ring resonators and subsequently generalize the construction to an arbitrary number $N$ of resonators. Each directional coupler is described by the standard relation
\begin{equation}
    \begin{pmatrix}
        b^{(out)}_1 \\
        b^{(out)}_2
    \end{pmatrix}=
\begin{pmatrix}
t&i\kappa\\i\kappa&t
\end{pmatrix} 
    \begin{pmatrix}
        a^{(in)}_1 \\
        a^{(in)}_2
    \end{pmatrix},\label{eq:coupler}
\end{equation}
where $t$ and $\kappa\in \mathbb{R}$ are the through- and cross-coupling amplitudes, respectively.

\begin{figure}
    \centering
    \includegraphics[width=0.5\textwidth]{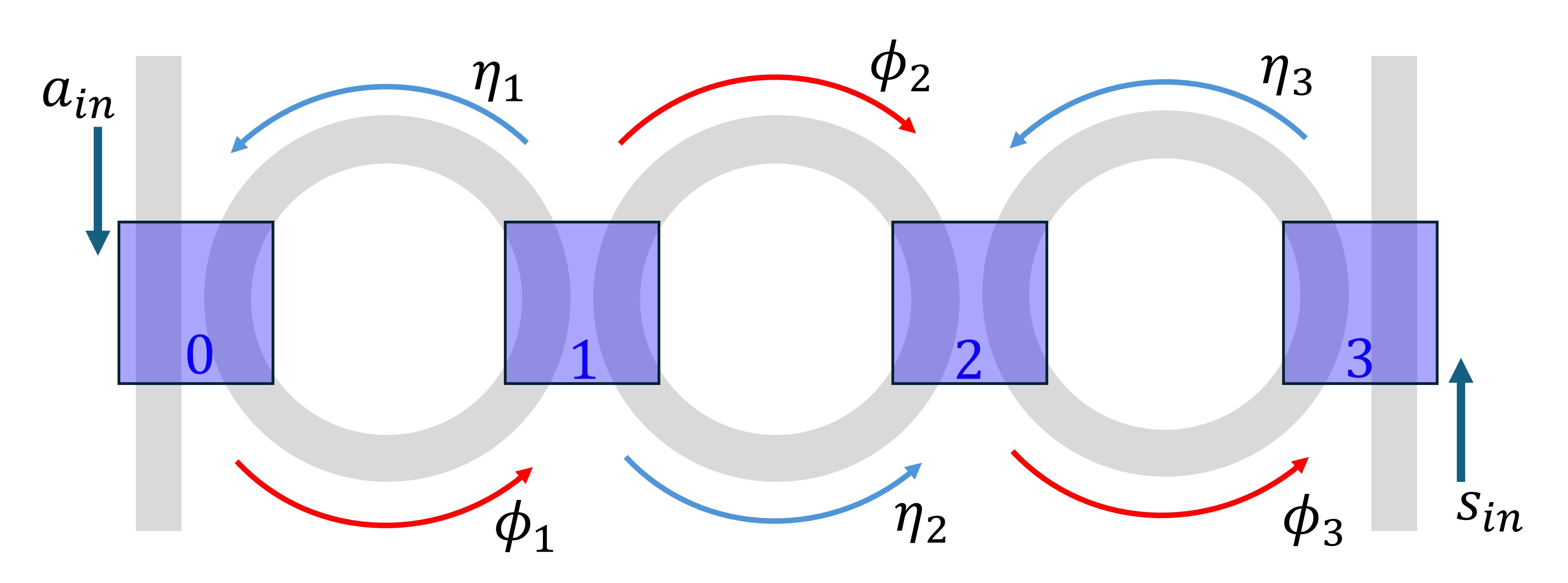}
    \caption{Schematics of a three-ring-resonators chain. The input fields $a_{in}$ and $s_{in}$ at the ports 
    excite internal fields $\phi_n$ and $\eta_n$ at resonator $n=1,2,3$. 
    Each directional coupler (blue boxes) determines the wave propagation via Eq. (\ref{eq:coupler}).}
    \label{fig:three_rings}
\end{figure}

For three resonators, we label the four couplers by $j=0,\ldots,3$.
The pairs $(t_0,\kappa_0)$ and $(t_3,\kappa_3)$ describe the input and
output couplers, respectively, whereas $j=1,2$ label the two
inter-resonator couplers. Propagation along half of resonator $n$ is
described by the complex propagation factor
$P_n\equiv\exp(i\theta_n/2)$, where
$\theta_n=(\omega/c)n_{\rm eff}L+i\alpha L/2$ is the complex phase
accumulated over half a ring trip (of length $L$). Its real part accounts for
the propagation phase, while its imaginary part accounts for
attenuation or amplification. In our simulations we chose $\alpha=0$.

Here we adopt a reduced description where we retain only one field amplitude at each resonator $n$, denoted by $\phi_n$ (see Fig. \ref{fig:three_rings}). The complementary amplitudes $\eta_n$ can be obtained from the $\phi_n$, and thus will be eliminated. Owing to the geometry of the chain, the physical propagation direction associated with these amplitudes alternates between clockwise and counterclockwise in successive rings.

The input and output boundaries satisfy
$\phi_1=i\kappa_0a_{\rm in}+t_0P_1\eta_1$ and
$\eta_3=t_3P_3\phi_3+i\kappa_3s_{\rm in}$.
At an internal coupler $n=1,2$, the corresponding relations are
$\eta_n=t_nP_n\phi_n+i\kappa_nP_{n+1}\eta_{n+1}$ and
$\phi_{n+1}=i\kappa_nP_n\phi_n+t_nP_{n+1}\eta_{n+1}$.
Eliminating $\eta_1,\eta_2,\eta_3$ exactly gives
\begin{equation}
\mathbf A_3\boldsymbol{\phi}=\mathbf D_3\mathbf s^+,
\quad
\boldsymbol{\phi}=
\begin{pmatrix}\phi_1\\\phi_2\\\phi_3\end{pmatrix},
\quad
\mathbf s^+=
\begin{pmatrix}a_{\rm in}\\s_{\rm in}\end{pmatrix},
\label{eq:three_ring_system}
\end{equation}
with
\begin{equation}
\mathbf A_3=
\begin{pmatrix}
t_0t_1P_1^2-1 &
i t_0\kappa_1t_2P_1P_2^2 &
-t_0\kappa_1\kappa_2t_3P_1P_2P_3^2
\\
i\kappa_1P_1 &
t_1t_2P_2^2-1 &
i t_1\kappa_2t_3P_2P_3^2
\\
0 &
i\kappa_2P_2 &
t_2t_3P_3^2-1
\end{pmatrix},\nonumber
\end{equation}
and
\begin{equation}
\mathbf D_3=
\begin{pmatrix}
-i\kappa_0 &
i t_0\kappa_1\kappa_2\kappa_3P_1P_2P_3
\\
0 &
t_1\kappa_2\kappa_3P_2P_3
\\
0 &
-i t_2\kappa_3P_3
\end{pmatrix}.\nonumber
\end{equation}
An important feature of the exact reduction is the appearance of effective longer-range matrix elements. For example, $A_{13}$ is proportional to $\kappa_1\kappa_2$, although the original optical network contains only nearest-neighbor couplers. Such terms are generated by eliminating the auxiliary fields $\eta_n$ and therefore represent multiple successive cross-coupling events rather than direct physical couplings between nonadjacent resonators.

We now consider a chain of $N$ resonators bounded by $N+1$ couplers, labeled $j=0,\ldots,N$. Resonator $n$ lies between couplers $n-1$ and $n$. For every internal coupler, $n=1,\ldots,N-1$, the same local relations apply,
$\eta_n=t_nP_n\phi_n+i\kappa_nP_{n+1}\eta_{n+1}$ and
$\phi_{n+1}=i\kappa_nP_n\phi_n+t_nP_{n+1}\eta_{n+1}$,
with boundary conditions
$\phi_1=i\kappa_0a_{\rm in}+t_0P_1\eta_1$ and
$\eta_N=t_NP_N\phi_N+i\kappa_Ns_{\rm in}$.
Recursive elimination of $\eta_N,\eta_{N-1},\ldots,\eta_1$ yields again
$\mathbf A\boldsymbol{\phi}=\mathbf D\mathbf s^+$, now with
$\boldsymbol{\phi}=(\phi_1,\ldots,\phi_N)^T$.
The exact matrix elements can be written compactly as
\begin{equation}
A_{nm}=
\begin{cases}
0, & m<n-1,\\[2pt]
i\kappa_{n-1}P_{n-1}, & m=n-1,\\[2pt]
i^{\,m-n}t_{n-1}t_m P_m^2
\displaystyle\prod_{j=n}^{m-1}(\kappa_j P_j)
-\delta_{nm}, & m\ge n .
\end{cases}
\label{eq:A_exact}
\end{equation}
where an empty product is understood as unity. The source matrix has only a left-incidence contribution in its first row,
$D_{n1}=-i\kappa_0\delta_{n1}$, whereas right incidence generates
$D_{n2}=-i^{N-n+1}t_{n-1}
(\prod_{j=n}^{N}\kappa_j)
(\prod_{j=n}^{N}P_j)$.

Equation~\eqref{eq:A_exact} shows that the exact reduced operator is upper Hessenberg rather than tridiagonal. In particular, an element $A_{n,n+p}$ contains $p$ consecutive cross-coupling amplitudes and is therefore of order
$\prod_{j=n}^{n+p-1}\kappa_j$. This provides a systematic hierarchy that we can take advantage of in the case of a weak-coupling approximation. 

Neglecting products containing two or more cross-coupling amplitudes, while retaining the through-coupling coefficients $t_n$ explicitly, leaves only the diagonal and nearest-neighbor elements,
\begin{eqnarray}
A_{nn}&=&t_{n-1}t_nP_n^2-1, \nonumber \\
A_{n,n-1}&=&i\kappa_{n-1}P_{n-1}, \label{eq:A_weak}\\
A_{n,n+1}&=&i t_{n-1}\kappa_nt_{n+1}P_nP_{n+1}^2 \nonumber,
\end{eqnarray}
with $D_{11}=-i\kappa_0$ and $D_{N2}=-i t_{N-1}\kappa_NP_N$ as the only source terms to the same order. The reduced problem therefore becomes tridiagonal and contains only effective nearest-neighbor interactions.

\subsection{General similarity transformations and experimental accessibility}
\label{app:general_similarity}

In the main text, we considered diagonal mappings which imposed restrictions on the network couplings. Here we address the question of whether more general 
similarity transformations can extend to systems that do not satisfy these restrictions. We show that such an extension is possible with complete channel 
access or a suitable structural symmetry. Ultimately, a practical \textit{in-situ} implementation of this scheme requires a fixed similarity transformation. When the on-site resonant frequencies are used as training parameters, the corresponding transformations are diagonal, turning the reconstruction of the adjoint field 
into a simple rescaling.

We define the fixed configuration system operator $\mathbf A=\omega\mathbf I-\mathbf H_0+\frac{i}{2}\mathbf D^T\mathbf D$ and the adjoint source $\mathbf q=(\partial g/\partial\boldsymbol\phi)^T$ associated with the objective $g$. The adjoint field $\boldsymbol\lambda$ satisfies $\mathbf A^T\boldsymbol\lambda=\mathbf q$. If a known $\mathbf\Xi$ satisfies the similarity transformation $\mathbf A^T=\mathbf\Xi^{-1}\mathbf A\mathbf\Xi$, then the adjoint field can be measured by an auxiliary experiment on the original device,
\begin{subequations}
\label{eq:app_general_drive}
\begin{align}
\mathbf A\widetilde{\boldsymbol\lambda}
&=i\mathbf D^T\widetilde{\mathbf S}^{+}
=\widetilde{\mathbf b}\equiv\mathbf\Xi\mathbf q,\\
\widetilde{\mathbf S}^{-}
&=-\widetilde{\mathbf S}^{+}+\mathbf D\widetilde{\boldsymbol\lambda},\\
\boldsymbol\lambda
&=\mathbf\Xi^{-1}\widetilde{\boldsymbol\lambda},
\end{align}
\end{subequations}
provided that $\widetilde{\mathbf b}$ can be generated through the available channels and the components of $\widetilde{\boldsymbol\lambda}$ entering the gradient can be measured. A transformation satisfying $\mathbf H_0^T=\mathbf\Xi^{-1}\mathbf H_0\mathbf\Xi$ also applies to $\mathbf A$ if $\mathbf\Xi$ commutes with $\mathbf D^T\mathbf D$. We first consider complete channel access, i.e., the capability of measuring the local fields and injecting excitations at every site of the network. Such knowledge permits dense transformations, and then structural mappings compatible with partial access. Finally, we show that a transformation which remains fixed under tuning of every onsite terms must be diagonal.

\subsubsection{Complete channel access}
\label{app:complete_channel_access}

With one channel per site, $\mathbf D=\operatorname{diag}(d_1,\ldots,d_N)$ with $d_j>0$, any adjoint source can be constructed and the full auxiliary field $\widetilde{\boldsymbol\lambda}$ can be measured. Therefore, Eq.~\eqref{eq:app_general_drive} gives
\begin{subequations}
\label{eq:app_full_access}
\begin{align}
\widetilde{\mathbf S}^{+}
&=-i\mathbf D^{-1}\mathbf\Xi\mathbf q,\\
\boldsymbol\lambda
&=\mathbf\Xi^{-1}\mathbf D^{-1}
\left(\widetilde{\mathbf S}^{-}+\widetilde{\mathbf S}^{+}\right),
\end{align}
\end{subequations}
so a dense $\mathbf\Xi$ is guaranteed to exist and $\widetilde{\boldsymbol\lambda}$ can be implemented by coherent multichannel excitation. Since every square matrix is similar to its transpose~\cite{taussky1959similarity}, complete access removes all restrictions on the coupling ratios, including networks that violate the cycle condition. For example, if $\mathbf A=\mathbf V\boldsymbol\Lambda\mathbf V^{-1}$ is diagonalizable, the transformation can be constructed by solving the corresponding eigenvalue problem. Then  $\mathbf\Xi=\mathbf V\mathbf V^T$ satisfies $\mathbf A^T=\mathbf\Xi^{-1}\mathbf A\mathbf\Xi$. However, efficient \textit{in-situ} implementation protocols benefit from avoiding transformations $\mathbf{\Xi}$ that depend on the tunable parameters as these must be solved and updated after each step (see Sec.~\ref{app:fixed_onsite_similarity}).

\subsubsection{Structural symmetry with partial access}
\label{app:structural_symmetry}

When only some sites have channels, a general dense mapping may transform an accessible adjoint source into one that requires excitation at internal inaccessible sites. A sufficient way to avoid this obstruction is to use a mapping that does not mix accessible and inaccessible sites. To understand this issue, we order the sites into an accessible set $L$ and an inaccessible set $I$, and define
\begin{equation}
\mathbf D=(\mathbf D_L\ \mathbf 0),
\qquad
\mathbf\Xi=
\begin{pmatrix}
\mathbf\Xi_L & \mathbf 0\\
\mathbf 0 & \mathbf\Xi_I
\end{pmatrix},
\end{equation}
where $\mathbf D_L$ is diagonal and nonsingular. Assume that the objective depends only on fields at accessible sites, such that $\mathbf q=(\mathbf q_L^T,\mathbf 0^T)^T$. Then the transformed source $\mathbf\Xi\mathbf q$ is also supported on $L$. If $\mathbf A^T=\mathbf\Xi^{-1}\mathbf A\mathbf\Xi$, the auxiliary excitation and accessible adjoint field are therefore
\begin{subequations}
\label{eq:app_block_reconstruction}
\begin{align}
\widetilde{\mathbf S}^{+}
&=-i\mathbf D_L^{-1}\mathbf\Xi_L\mathbf q_L,\\
\boldsymbol\lambda
&=\boldsymbol\lambda_L=\mathbf\Xi_L^{-1}\mathbf D_L^{-1}
\left(\widetilde{\mathbf S}^{-}+\widetilde{\mathbf S}^{+}\right).
\end{align}
\end{subequations}
Thus, the adjoint field can be reconstructed using only channel measurements, even when the mapping is not diagonal. The similarity relation must hold for the entire wave operator, including the couplings between accessible and inaccessible sites. One possible construction permutes the sites such that the resulting operator equals $\mathbf A^T$. The permutation must keep accessible sites within $L$ and internal sites within $I$.

Let $\sigma(i)$ denote the relabeled site corresponding to site $i$. The required conditions are $\sigma(L)=L$, $\sigma(I)=I$, and
\begin{equation}
A_{\sigma(i)\sigma(j)}=A_{ji}
\qquad\text{for all }i,j.
\end{equation}
Thus, the relabeling of each pair of sites must reproduce the matrix entry with its indices reversed. The corresponding permutation matrix, which plays the role of the similarity transformation in this construction, $\Xi_{mn}=\delta_{m,\sigma(n)}$, then satisfies $\mathbf\Xi^{-1}\mathbf A\mathbf\Xi=\mathbf A^T$.

These permutation based transformations can exist even in networks that admit no diagonal mapping. However, a fixed permutation remains valid only while tuning preserves the relation above. Physically, this sites exchanged by the permutation must have the same local resonance frequency and total decay rate, and the permutation must map each directed coupling onto an equal coupling in the reverse direction between the corresponding sites. This means that the resonance frequencies and intrinsic decay rates of the sites must therefore be tuned together rather than independently, as well as the permuted couplings, imposing restrictions on the allowed optimization parameters.

\subsubsection{Fixed mappings under independent onsite tuning}
\label{app:fixed_onsite_similarity}
We now ask when a mapping can remain fixed under independent tuning of the onsite potentials, in contrast to the permuted transformation described above. Therefore, we keep the couplings fixed, and write
\begin{equation}
\mathbf A(\boldsymbol\theta)
=\mathbf A_0+\sum_j\theta_j\mathbf E_{jj},
\end{equation}
where $\mathbf E_{jj}$ has a single nonzero entry, equal to one, at position $(j,j)$, and $\theta_j$ is the onsite shift at site $j$.

We suppose a single nonsingular similarity transformation $\mathbf\Xi$ satisfies
\begin{equation}
\mathbf A(\boldsymbol\theta)\mathbf\Xi
=\mathbf\Xi\mathbf A(\boldsymbol\theta)^T
\end{equation}
for every allowed parameter setting.
Subtracting this relation at two settings that differ only in $\theta_j$ gives
\begin{equation}
\mathbf E_{jj}\mathbf\Xi=\mathbf\Xi\mathbf E_{jj}.
\end{equation}
Comparing the $(j,k)$ and $(k,j)$ entries immediately yields $\Xi_{jk}=\Xi_{kj}=0$ for every $k\neq j$. Thus, if every onsite term can be varied independently, $\mathbf\Xi$ must be diagonal. If independent onsite tuning is available only at certain sites, the off-diagonal entries in the rows and columns associated with those sites must vanish.

Conversely, a nonsingular diagonal matrix $\mathbf\Xi$ that satisfies $\mathbf A_0\mathbf\Xi=\mathbf\Xi\mathbf A_0^T$ commutes with every onsite shift and therefore remains valid throughout onsite tuning. Thus, among mappings that satisfy the similarity relation for $\mathbf A_0$, only the diagonal mappings remain valid under arbitrary independent tuning of all onsite terms.

\end{document}